\documentstyle[12pt,fleqn]{article} \textheight=9in
\newcommand{\ra}{\rightarrow}
\newcommand{\bra}{\langle} \newcommand{\ket}{\rangle}
\newcommand{\be}{\begin{equation}}
\newcommand{\ee}{\end{equation}}
\newcommand{\bea}{\begin{eqnarray}}
\newcommand{\eea}{\end{eqnarray}}

\newcommand{\E}{\mbox{e}}
\newcommand{\e}{\mbox{\scriptsize e}}
\newcommand{\ffi}{\varphi}

\newcommand{\ep}{\qquad {\vrule height 10pt width 8pt depth 0pt}}

\newcommand{\grintl}{[\kern-.18em [}
\newcommand{\grintr}{]\kern-.18em ]}
\newcommand{\ds}{\displaystyle}
\newtheorem{lem}{Lemma}[section]

\newtheorem{thm}{Theorem}[section]

\def\R{\hbox{$\mit I$\kern-.277em$\mit R$}}
\def\C{\hbox{$\mit I$\kern-.6em$\mit C$}}
\def\un{\hbox{$\mit I$\kern-.77em$\mit I$}}
\def\0{\hbox{$\mit I$\kern-.70em$\mit O$}}
\def\r{I\kern-.277em R}

\begin{document}

\title{Semiclassical Dynamics with Exponentially Small
Error Estimates}
\author{George A. Hagedorn\thanks{Partially
Supported by National Science Foundation
Grant DMS--9703751.}\\
Department of Mathematics and\\
Center for Statistical Mechanics and Mathematical Physics\\
Virginia Polytechnic Institute and State University\\
Blacksburg, Virginia 24061-0123, U.S.A.\\[15pt]
\and
Alain Joye\thanks{Partially Supported by
Fonds National
Suisse de la Recherche Scientifique, Grant 8220-037200.}\\
Institut Fourier\\ Unit\'e Mixte de Recherche CNRS-UJF 5582\\
Universit\'e de Grenoble I\\
BP 74\\
F--38402 Saint Martin d'H\`eres Cedex, France}
\date{}
\maketitle

\begin{abstract}
We construct approximate solutions to the time--dependent Schr\"odinger equation
$$
i\,\hbar\,\frac{\partial \psi}{\partial t}\ =\
-\,\frac{\hbar^2}{2}\,\Delta\,\psi\,+\,V\,\psi
$$
for small values of $\hbar$.
If $V$ satisfies appropriate analyticity and growth hypotheses and $|t|\le T$,
these solutions agree with exact solutions up to errors whose norms are
bounded by
$$
C\ \exp\left\{\,-\,\gamma/\hbar\,\right\} ,
$$
for some $C$ and $\gamma>0$.
Under more restrictive hypotheses, we prove that for sufficiently small $T'$,
$|t|\le T'\,|\log(\hbar)|$ implies the norms of the errors are bounded by
$$
C'\ \exp\left\{\,-\,\gamma'/\hbar^{\sigma}\,\right\} ,
$$
for some $C'$, $\gamma'>0$, and $\sigma>0$.

\end{abstract}

\newpage

\section{Introduction}

In this paper, we construct exponentially accurate semiclassical
approximations $\psi(x,\,t,\,\hbar)$ to certain normalized
exact solutions $\Psi(x,\,t,\,\hbar)$ of the
$d$--dimensional time--dependent Schr\"od\-ing\-er equation
\be\label{sch}
i\,\hbar\,\frac{\partial\Psi}{\partial t}\ =\
-\,\frac{\hbar^2}{2}\,\Delta\,\Psi\,+\,V\,\Psi .
\ee
More precisely, our main result is that for $|t|\le T$ and small values of
$\hbar$, these approximations satisfy error estimates of the form
\be\label{expest}
\|\,\psi(x,\,t,\,\hbar)\,-\,\Psi(x,\,t,\,\hbar)\,\|_{L^2(\r^d)}
\ \le\ \,C\ \exp\left\{\,-\,\gamma/\hbar\,\right\} ,
\ee
where $\gamma >0$.

Our construction of $\psi(x,\,t,\,\hbar)$ is technically complicated, but quite
explicit. It uses a particular collection of semiclassical wave packets
$\{\,\ffi_j(A,\,B,\,\hbar,\,a,\,\eta,\,\cdot\,)\,\}$ that are defined in
\cite{semi4}, \cite{raise}, and the next section. Here $A$ and $B$ are
$d\times d$ complex matrices that satisfy certain conditions. The quantities $a$
and $\eta$ are elements of $\R^d$. For fixed $A$, $B$, $\hbar$, $a$, and $\eta$,
$\{\,\ffi_j(A,\,B,\,\hbar,\,a,\,\eta,\,\cdot\,)\,\}$ is an orthonormal basis
of $L^2(\R^d)$ as $j$ ranges over all $d$-dimensional multi-indices.

The function $\ffi_j(A,\,B,\,\hbar,\,a,\,\eta,\,\cdot\,)$ is concentrated
near position $a$, and its Fourier transform is concentrated near momentum
$\eta$. Its position and momentum uncertainties are proportional to
$\sqrt{\hbar}$. The position uncertainty is determined by the matrix
$|A|=\sqrt{A\,A^*}$, and the momentum uncertainty is determined by
$|B|=\sqrt{B\,B^*}$.

We construct  $\psi(x,\,t,\,\hbar)$ by applying the idea of ``optimal
truncation'' of an asymptotic expansion.
For initial conditions of the form
\be\label{incon}
\Psi(x,\,0,\,\hbar)\,=\,\sum_{|j|\le J}\,c_j\,
\ffi_j(A(0),B(0),\hbar,a(0),\eta(0),x),
\ee
with $\sum_{|j|\le J}\,|c_j|^2\,=\,1$,
there exist (\cite{semi3}, \cite{semi4}) approximate solutions
\be\label{app}
\psi_l(x,\,t,\,\hbar)\,=\,e^{iS(t)/\hbar}\
\sum_{|j|\le \widetilde{J}(l)}\,
c_j(l,\,t,\,\hbar)\,\ffi_j(A(t),B(t),\hbar,a(t),\eta(t),x),
\ee
that satisfy
\be\label{est}
\sup_{t\in [-T,T]}\ \|\psi_l(x,\,t,\,\hbar)-\Psi(x,\,t,\,\hbar)\|_{L^2(\r^d)}
\,\leq\,C(l)\,h^{l/2}
\ee
for some constant $C(l)$.
Here $\widetilde{J}(l)=J+3l-3$, and $A(t)$, $B(t)$, $a(t)$, $\eta(t)$, and
$S(t)$ are solutions to the classical equations of motion
\bea
\dot{a}(t)&=&\eta(t) ,\nonumber \\
\dot{\eta}(t)&=&-\,\nabla V(a(t)),\nonumber \\
\dot{A}(t)&=&i\,B(t),\label{newton} \\
\dot{B}(t)&=&i\,V^{(2)}(a(t))\,A(t),\nonumber \\
\dot{S}(t)&=&\frac{\eta(t)^2}2\,-\,V(a(t)),\nonumber
\eea
where $V^{(2)}$ denotes the Hessian matrix for $V$, and the
initial conditions $A(0)$, $B(0)$, $a(0)$, $\eta(0)$, and
$S(0)=0$ satisfy
\bea\label{initcond}
A^t(0)\,B(0)\,-\,B^t(0)\,A(0)&=&0,\\
A^*(0)\,B(0)\,+\,B^*(0)\,A(0)&=&2\,I.
\eea
The $c_j(l,\,t,\,\hbar)$ satisfy a linear system of ordinary differential
equations that we describe in the next section.

We carefully estimate the $l$--dependence of $C(l)$ in (\ref{est}).
Then, for each $\hbar$, we choose $l(\hbar)$ to minimize the error
$C(l(\hbar))\,\hbar^{l(\hbar)/2}$ over all choices of $l$. It
turns out that $l(\hbar)$ behaves like a constant times $1/\hbar$. We define
$\psi(x,\,t,\,\hbar)=\psi_{l(\hbar)}(x,\,t,\,\hbar)$ and
prove that (\ref{expest}) is satisfied.

For $t$ in a fixed compact interval, the precise statement of our results is the
following:

\vskip .25cm
\noindent
\begin{thm}\label{mainthm}
Suppose $V$ is a real-valued function on $\R^d$ that is bounded below and
has an analytic continuation to the set
\be\nonumber
D\,=\,\{\,z\in\C^d\,:\ |\mbox{Im }z_j|\,<\,\delta,\ j=1,\,2,\,\cdots,\,d\,\} .
\ee
Suppose further that there exist $M>0$ and $\tau>0$, such that
\be\nonumber
|\,V(z)\,|\,\leq\,M\,\exp(\tau |z|^2),\qquad \mbox{for all } z\in D,
\ee
where $|z|^2\,=\,\sum_{j=1}^d\,|z_j|^2$.
Suppose initial conditions $A(0)$, $B(0)$, $a(0)$, $\eta(0)$, $S(0)=0$, and
$c_j$ for $|j|\le J$ are specified that satisfy (\ref{initcond}) and
$\sum_{|j|\le J}\,|c_j|^2\,=\,1$.
Then for any $T>0$, there exist $C$ and $\gamma >0$,
such that the difference between the semiclassical
approximation $\psi(x,\,t,\,\hbar)$ and the exact solution $\Psi(x,\,t,\,\hbar)$
to the Schr\"odinger equation (\ref{sch}) with initial condition
(\ref{incon}) satisfies
$$
\|\,\psi(x,\,t,\,\hbar)\,-\,\Psi(x,\,t,\,\hbar)\,\|_{L^2(\r^d)}
\ \le\ \,C\ \exp\left\{\,-\,\gamma/\hbar\,\right\} ,
$$
whenever $|t|\,\le\,T$.
\end{thm}

\vskip .25cm
\noindent
{\bf Remarks}\quad 1.\quad Theorem \ref{mainthm} can be generalized to allow
time--dependent potentials. For example, suppose a potential $V(x,\,t)$
depends smoothly on $t$, is bounded below, and satisfies
$$
\frac{1}{m!}\,|D^mV(x,\,t))|\ \leq\
\frac{M\,\exp(\tau |x|^2)}{\delta^{|m|}},
$$
for $|t|\le T$ for all multi-indices $m$.
Suppose further that a classical solution $(a(t),\,\eta(t))$ to
Newton's equations with potential $V(x,\,t)$ is bounded for $|t|\le T$. Then the
conclusion to Theorem \ref{mainthm} holds.\\[7pt]
2.\quad Our results can also be extended to obtain weaker error estimates of the
form\linebreak $C\ \exp\left\{\,-\,\gamma/\hbar^{\sigma}\,\right\}$ for some
$\sigma\in (0,\,1)$, when the potential belongs to a Gevrey class.\\[7pt]
3.\quad Theorem \ref{mainthm} is optimal in the sense that the conclusion fails
if the hypotheses are relaxed slightly.
For example, consider the one-dimensional potential
$V(x)=\exp (-1/x^u)$, for $x>0$, and $V(x)=0$ for $x\leq 0$,
where $u>0$. It is shown in \cite{Jung} that this potential belongs to the
Gevrey class of order $1+1/u$.
For initial conditions $a(0)=0$, $\eta(0)=0$,
$A(0)=B(0)=1$, $S(0)=0$, and $c_j(0)=\delta_{j,0}$, our approximation yields
$\ffi_0(1+it, 1, \hbar, 0, 0)$, for all times. This function
is very simple, and we can write the error term explicitly.
By steepest descent analysis, we can show that there exist $\delta>0$ and
$\Sigma_1>\Sigma_2>0$, such that
$t\in (0,\hbar^{\delta})$ implies
\bea\nonumber
\exp(-\Sigma_1/\hbar^{u/(1+u)})
&\le&\|\,e^{-itH(\hbar)/\hbar}\,\ffi_0(1,1,\hbar,0,0,\cdot)
\,-\,\ffi_0(1+it,1,\hbar,0,0,\cdot)\,\|\\[6pt] \nonumber
&\le&\exp(-\Sigma_2/\hbar^{u/(1+u)}).
\eea
Note also that if we choose $a(0)=-a<0$ and $ \eta(0)=\eta>0$, it is
easy to check that the error term is $O(\exp(-\gamma /\hbar))$,
for each $t<a/\eta$.\\[7pt]
4.\quad For all practical purposes, we can replace the $c_j$'s by
the corresponding Dyson expansion up to order $l(\hbar)$ without
spoiling our exponential estimate. The normalization of the
approximation, however, will be lost.

\vskip .25cm
Theorem \ref{mainthm} can also be generalized to allow time intervals that grow
like $|\log(\hbar)|$ as $\hbar$ tends to zero. However, we obtain a somewhat
weaker conclusion. Our precise results are summarized by the following theorem.

\vskip .25cm
\noindent
\begin{thm}\label{longtime}
Suppose $V$ is bounded below and analytic in
\be\nonumber
D\,=\,\{\,z\in\C^d\,:\ |\mbox{Im }z_j|\,<\,\delta,\ j=1,\,2,\,\cdots,\,d\,\} .
\ee
Suppose further that there exist $M>0$ and $\tau>0$, such that
\be\nonumber
|\,V(z)\,|\,\leq\,M\,\exp(\tau |z|),\qquad \mbox{for all } z\in D.
\ee
Suppose initial conditions $A(0)$, $B(0)$, $a(0)$, $\eta(0)$, $S(0)=0$, and
$c_j$ for $|j|\le J$ are specified that satisfy (\ref{initcond}) and
$\sum_{|j|\le J}\,|c_j|^2\,=\,1$, and further assume there exist $N>0$
and$\lambda >0$, such that
\be\label{expgro}
\|A(t)\|\,\leq\,N\,\exp (\lambda |t|).
\ee
Then for sufficiently small $T'>0$, there exist $C'$, $\gamma'>0$, and
$\sigma>0$, such that the difference between the semiclassical
approximation $\psi(x,\,t,\,\hbar)$ and the exact solution $\Psi(x,\,t,\,\hbar)$
to the Schr\"odinger equation (\ref{sch}) with initial condition
(\ref{incon}) satisfies
$$
\|\,\psi(x,\,t,\,\hbar)\,-\,\Psi(x,\,t,\,\hbar)\,\|_{L^2(\r^d)}
\ \le\ \,C'\ \exp\left\{\,-\,\gamma'/\hbar^{\sigma}\,\right\} ,
$$
whenever $|t|\,\le\,T'\,|\log(\hbar)|$.
\end{thm}

\vskip .25cm
\noindent
{\bf Remark}\quad
Standard existence and uniqueness theorems for systems of ODE's show that
condition (\ref{expgro}) is satisfied if
the norm of the Hessian $V^{(2)}(a(t))$ is uniformly bounded.
That is the
case if $V$ is the sum of a quadratic polynomial plus a bounded analytic
function. It is also the case if
$E$ denotes the energy of the considered trajectory and
$\|V^{(2)}(x)\|$ is bounded on the connected
component of the classically allowed region $D_E=\{ x\in\R^d\,:\,V(x)\leq E\}$
that contains $a(t)$.
This is satisfied for all confining potentials.

\vskip .25cm
The propagation of coherent states is also considered by Combescure
and Robert in \cite{CR}, using an approximation given by a
linear combination of squeezed states. (The squeezed states coincide with our
semiclassical wave packets, although the notation is quite different.)
Their emphasis is on the long time behavior of this approximation. The bound
on the error term is of the form $C_{l}(t)\hbar^{l/2}$, with expicit control
of the time-dependence of $C_{l}(t)$ in terms of classical quantities. The
$l$ behavior is however not investigated.

Results of a flavor similar to ours can be found in the work of
Yajima \cite{Yajima}. They are obtained by means
of the pseudo-differential techniques developed in the analytic
context by Sj\"ostrand in \cite{Sjostrand}. These results concern the
propagation of wave packets of
the form $\ffi(x)=\E^{iS(x)/\hbar}f(x)$, where $S$ is analytic and
$f$ belongs to the set
of compactly supported Gevrey functions of order $s>1$.
Assuming the potential $V$ is analytic, Yajima
constructs approximations to the actual evolution of
such wave packets that are valid up to an error, whose $L^2(\R^d)$ norm
is of order $\E^{-\gamma/\hbar^{1/(2s-1)}}$, with $\gamma >0$ (see
Theorems 1.2, 1.2 and Lemma 2.5 in \cite{Yajima}). However, it should be
possible to make use of the theory \cite{Sjostrand} to recover our results.

Similar issues have been dealt with by Bambusi, Graffi and Paul in
\cite{BGP}. They focus on the validity for large times of the semiclassical
approximation of the Heisenberg evolution of a smooth observable,
under analyticity assumptions on the hamiltonian. They prove that the
semiclassical approximation remains useful for times up to order
$|\ln(\hbar)|$, the Ehrensfest time scale. However, the Hamiltonians
they can accomodate consist more or less of analytic perturbations of
the harmonic oscillator that decay as $x$ and $p$ tend to infinity.

\vskip .25cm
The paper is organized as follows: In the Section 2, we prove Theorem
\ref{mainthm} under the assumption that two types of error terms satisfy
certain bounds. We prove the two required bounds in Sections 3 and 4.
In Section 5 we describe the proof of Theorem \ref{longtime}.

\vskip .5cm
\section{Proof of Theorem 1.1}
\setcounter{equation}{0}

We begin this section by presenting the definition of the semiclassical wave
packets\\
$\ffi_j(A,\,B,\,\hbar,\,a,\,\eta,\,x)$ that is given in \cite{raise}.
A more explicit, but more complicated definition is given in \cite{semi4}.
Since \cite{raise} provides a detailed discussion of these wave packets, we
do not prove all their properties here.

We adopt the standard multi-index notation. A multi-index
$j=(j_1,\,j_2,\,\dots ,\,j_d)$ is a $d$-tuple of non-negative integers.
We define
$|j|=\sum_{k=1}^d\,j_k,\quad
j!=(j_1!)(j_2!)\cdots(j_d!)$,\\
$x^j=x_1^{j_1}x_2^{j_2}\cdots x_d^{j_d},\quad\mbox{and}\quad
\ds D^j=\frac{\partial^{|j|}}{(\partial x_1)^{j_1}(\partial x_2)^{j_2}\cdots
(\partial x_d)^{j_d}}$.

Throughout the paper we assume $a\in\R^d$, $\eta\in\R^d$ and $\hbar>0$. We
also assume that $A$ and $B$ are $d\times d$ complex invertible matrices that
satisfy
\bea\label{cond1}
A^t\,B\,-\,B^t\,A&=&0,\\ \label{cond2}
A^*\,B\,+\,B^*\,A&=&2\,I.
\eea
These conditions guarantee that both the real and imaginary parts of $BA^{-1}$
are symmetric. Furthermore, $\mbox{Re}\,BA^{-1}$ is strictly positive definite,
and $\left(\,\mbox{Re}\,BA^{-1}\,\right)^{-1}\,=\,A\,A^*$. We note that
conditions (\ref{cond1}) and (\ref{cond2}) are preserved under the dynamics
generated by (\ref{newton}).

Our definition of $\ffi_j(A,\,B,\,\hbar,\,a,\,\eta,\,x)$ is based on the
following raising operators that are defined for $m=1,\,2,\,\dots ,\,d$.
$$
{\cal A}_m(A,B,\hbar,a,\eta)^*
\ =\
\frac{1}{\sqrt{2\hbar}}\,\left[\,\sum_{n=1}^d\,\overline{B}_{n\,m}\,(x_n-a_n)
\ -\ \sum_{n=1}^d\,\overline{A}_{n\,m}\,
(-i\hbar\frac{\partial\phantom{x^n}}{\partial x_n}-\eta_n)\,\right] .
$$

\vskip .25cm
\noindent
{\bf Definition}\quad
For the multi-index $j=0$, we define the normalized complex Gaussian wave packet
(modulo the sign of a square root) by
\bea\nonumber
&&\ffi_0(A,\,B,\,\hbar,\,a,\,\eta,\,x)\,=\,\pi^{-d/4}\,
\hbar^{-d/4}\,(\det(A))^{-1/2}\\[6pt]
&&\qquad\qquad\times\
\exp\left\{\,-\,\langle\,(x-a),\,B\,A^{-1}\,(x-a)\,\rangle/(2\hbar)\,
+\,i\,\langle\,\eta,\,(x-a)\,\rangle/\hbar\,\right\} .
\eea
Then, for any non-zero multi-index $j$, we define
\bea\nonumber
\ffi_j(A,\,B,\,\hbar,\,a,\,\eta,\,\cdot\,)&=&
\frac{1}{\sqrt{j!}}\
\left(\,{\cal A}_1(A,B,\hbar,a,\eta)^*\right)^{j_1}\
\left(\,{\cal A}_2(A,B,\hbar,a,\eta)^*\right)^{j_2}\ \cdots \\[5pt]
\nonumber &&\qquad\qquad\times\
\left(\,{\cal A}_d(A,B,\hbar,a,\eta)^*\right)^{j_d}\
\ffi_0(A\,,B,\,\hbar,\,a,\,\eta,\,\cdot\,).
\eea

\vskip .25cm
\noindent
{\bf Remarks}\quad
1.\quad For $A=B=I$, $\hbar=1$, and $a=\eta=0$, the
$\ffi_j(A,\,B,\,\hbar,\,a,\,\eta,\,\cdot\,)$ are just the standard Harmonic
oscillator eigenstates with energies $|j|+d/2$.\\[7pt]
2.\quad For each $A$, $B$, $\hbar$, $a$, and $\eta$,
the set $\{\,\ffi_j(A,\,B,\,\hbar,\,a,\,\eta,\,\cdot\,)\,\}$
is an orthonormal basis for $L^2(\R^d)$.\\[7pt]
3.\quad The raising operators can also be given by another formula that was
omitted from \cite{raise} in the multi-dimensional case.
If we set
$$g(A,\,B,\,\hbar,\,a,\,x)\,=\,
\exp\left\{\,-\,\langle\,(x-a),\,
\left(BA^{-1}\right)^*\,(x-a)\,\rangle/(2\hbar)\,-\,i\,
\langle\,\eta,\,(x-a)\,\rangle/\hbar\,\right\} ,$$
then we have
$$\left(\,{\cal A}_m(A,B,\hbar,a,\eta)^*\,\psi\,\right)(x)\,=\,
-\,\sqrt{\frac{\hbar}2}\,\frac 1{g(A,\,B,\,\hbar,\,a,\,x)}\,
\sum_{n=1}^d\,\overline{A}_{n\,m}\,
\frac{\partial\phantom{x^n}}{\partial x_n}\,
\left(\,g(A,\,B,\,\hbar,\,a,\,x)\,\psi(x)\,\right) .$$
4.\quad In \cite{semi4}, the state $\ffi_j(A,\,B,\,\hbar,\,a,\,\eta,\,x)$ is
defined as a normalization factor times
$$
{\cal H}_j(A;\,\hbar^{-1/2}\,|A|^{-1}\,(x-a))\
\ffi_0(A,\,B,\,\hbar,\,a,\,\eta,\,x).
$$
Here ${\cal H}_j(A;\,y)$ is a
$|j|^{\mbox{\scriptsize th}}$
order polynomial in $y$ that depends on $A$ only through
$U_A$, where $A=|A|\,U_A$ is the polar decomposition of $A$.\\[7pt]
5.\quad By the argument on page 370 of \cite{semi4} or by scaling out the $|A|$
and $\hbar$ dependence and using Remark 3 above, one can show that
${\cal H}_j(A;\,y)\,\E^{-y^2/2}$ is an (unnormalized) eigenstate of the usual
Harmonic oscillator with energy $|j|+d/2$.\\[7pt]
6.\quad When the dimension $d$ is $1$, the position and momentum uncertainties
of the\newline
$\ffi_j(A,\,B,\,\hbar,\,a,\,\eta,\,\cdot\,)$ are
$\sqrt{(j+1/2)\hbar}\,|A|$ and $\sqrt{(j+1/2)\hbar}\,|B|$, respectively. In
higher dimensions, they are bounded by
$\sqrt{(|j|+d/2)\hbar}\,\|A\|$ and $\sqrt{(|j|+d/2)\hbar}\,\|B\|$,
respectively.\\[7pt]
7.\quad When we approximately solve the Schr\"odinger equation, the choice of
the sign of the square root in the definition of
$\ffi_0(A,\,B,\,\hbar,\,a,\,\eta,\,\cdot\,)$ is determined by continuity in $t$
after an arbitrary initial choice.

\vskip .25cm
The proof of the theorem depends on the following abstract lemma.

\vskip .25cm
\noindent
\begin{lem}\label{fundcalc}
Suppose $H(\hbar)$ is a family of self-adjoint
operators for $\hbar>0$. Suppose $\psi(t,\,\hbar)$ belongs to the domain
of $H(\hbar)$, is continuously differentiable in $t$, and approximately
solves the Schr\"odinger equation in the sense that
\be\label{xidef}
i\,\hbar\,\frac{\partial\psi}{\partial t}(t,\,\hbar)\ =\
H(\hbar)\,\psi(t,\,\hbar)\ +\ \xi(t,\,\hbar),
\ee
where $\xi(t,\,\hbar)$ satisfies
\be\label{xiest}
\|\,\xi(t,\,\hbar)\,\|\ \le \,\mu(t,\,\hbar).
\ee
Then, for $t>0$,
\be\label{ultimate}
\|\,\E^{-itH(\hbar)/\hbar}\,\psi (0,\,\hbar)\ -\ \psi(t,\,\hbar)\,\|\
\le\ \hbar^{-1}\ \int_0^{t}\,\mu(s,\,\hbar)\,ds.
\ee
The analogous statement holds for $t<0$.
\end{lem}

\vskip .25cm
\noindent {\bf Proof}:\quad Assume $t>0$; the proof for $t<0$ is similar.
By the unitarity of the propagator $\E^{-itH(\hbar)/\hbar }$
and the fundamental theorem of calculus, the quantity on the
left-hand side of (\ref{ultimate}) can be estimated as follows:
\bea\nonumber
&&\|\,\E^{-itH(\hbar)/\hbar}
\,\psi(0,\,\hbar)\ -\ \psi(t,\,\hbar)\,\|\\[5pt]
\nonumber
&=&\|\,\psi(0,\,\hbar)\ -\ \E^{itH(\hbar)/\hbar}
\,\psi(t,\,\hbar)\,\|\\[7pt]
\nonumber
&=&\|\,\int_0^t\ \frac{\partial\phantom{s}}{\partial s}\,\left(
\,\psi(0,\,\hbar)\ -\ \E^{isH(\hbar)/\hbar}
\,\psi(s,\,\hbar)\,\right)\ ds\ \| \\[7pt]
\nonumber
&=&\|\,\int_0^t\ \left(\,-i\,\hbar^{-1}\,
\E^{isH(\hbar)/\hbar}\,H(\hbar)\,\psi(s,\,\hbar)\ -\
\E^{isH(\hbar)/\hbar}
\,\frac{\partial\psi}{\partial s}(s,\,\hbar)\,\right)\ ds\ \| \\[6pt]
\nonumber
&=&\|\,\int_0^t\ i\,\hbar^{-1}\,
\E^{isH(\hbar)/\hbar}\,\xi(s,\,\hbar)\ ds\ \|\\[5pt]
\nonumber
&\le&\hbar^{-1}\ \int_0^{t}\,
\mu(s,\hbar)\,ds.
\eea
This proves the lemma. \ep

\vskip .25cm
Because $V$ is smooth and bounded below, there exist global solutions to
the first two equations of the system (\ref{newton}) for any initial
condition. It then follows immediately that the remaining three
equations of the system (\ref{newton}) have global solutions.
Furthermore, it is not difficult (\cite{semi3}, \cite{semi4}) to prove
that (\ref{cond1}) and (\ref{cond2}) are preserved by the flow.

As mentioned in the introduction, it is proved in \cite{semi3} and \cite{semi4}
that initial conditions of the form (\ref{incon}) give rise to
approximate solutions of the form
$$
\psi_l(x,\,t,\,\hbar)\,=\,e^{iS(t)/\hbar}\
\sum_{|j|\le \widetilde{J}(l)}\,
c_j(l,\,t,\,\hbar)\,\ffi_j(A(t),B(t),\hbar,a(t),\eta(t),x),
$$
with errors whose norms are of order $\hbar^{l/2}$.
Here $\widetilde{J}(l)=J+3l-3$, and $A(t)$, $B(t)$, $a(t)$, $\eta(t)$, and
$S(t)$ satisfy (\ref{newton}). The coefficients $c_j(l,\,t,\,\hbar)$ satisfy the
linear system
\be\label{eqcj}
i\,\hbar\,\dot{c}_k(l,\,t,\,\hbar)\,=\,\sum_{|j|\le \widetilde{J}(l)}\,
K_{k\,j}(l,\,t,\,\hbar)\,c_j(l,\,t,\,\hbar),\qquad\qquad
|k|=0,\,1,\,\dots,\,\widetilde{J}(l),
\ee
with initial conditions $c_j(l,\,0,\,\hbar)=c_j$, for $|j|\le J$
in accordance with (\ref{incon}) and $c_j(l,\,0,\,\hbar)=0$ for $|j|>J$.

To specify the $(J+3l-2)\times (J+3l-2)$ matrix $K(l,\,t,\,\hbar)$
that appears in
(\ref{eqcj}), we first decompose the potential as
\be\label{defy}
V(x)=W_{a}(x)+Z_a(x)\equiv W_{a}(x)+(V(x)-W_a(x)),
\ee
where $W_a(x)$ denotes the second order Taylor
approximation (with the obvious abuse of notation)
\be
W_a(x)\,\equiv\,V(a)\,+\,V^{(1)}(a)\,(x-a)\,+\,V^{(2)}(a)\,(x-a)^2/2,
\ee
Next (reverting to multi-index notation), we approximate $Z_a(x)$ by its Taylor
approximation of order $l+1$,
\be\label{ypoly}
Z^{[l]}_a(x)\,=\,\sum_{3\le |m|\le l+1}\,\frac{(D^mV)(a)}{m!}\,(x-a)^m,
\ee
and define the infinite matrix
\bea\label{defk}
&&\widetilde{K}_{k\,j}(l,\,t,\,\hbar)\ =\nonumber \\[5pt]
&&\bra\,
\ffi_k(A(t),B(t),\hbar,a(t),\eta(t),x),\
Z^{[l]}_{a(t)}(x)\,\ffi_j(A(t),B(t),\hbar,a(t),\eta(t),x)\,\ket .
\eea
Then, we obtain the matrix $K(l,\,t,\,\hbar)$ from
$\widetilde{K}(l,\,t,\,\hbar)$ by restricting the indices to
$|j|\le\widetilde{J}(l)$ and $|k|\le\widetilde{J}(l)$.

The general strategy \cite{semi3}, \cite{semi4}, \cite{raise} to show that
$\psi_l(x,\,t,\,\hbar)$ is an approximation to the actual solution
$\Psi(x,\,t,\,\hbar)$ of (\ref{sch}) and (\ref{incon}) up to order
$\hbar^{l/2}$ is as follows:
From \cite{raise}, we know that for all multi-indices $j$,
\bea\label{profi}
&&i\,\hbar\,\frac{\partial }{\partial t}\,
\left[\,\E^{iS(t)/\hbar}\,\ffi_j(A(t),B(t),\hbar,a(t),\eta(t),x)\,\right]
\nonumber\\[8pt]
&=&\left(\,-\,\frac{\hbar^2}{2}\,\Delta\,+\,W_{a(t)}(x)\,\right)\,
\left[\,\E^{iS(t)/\hbar}\,\ffi_j(A(t),B(t),\hbar,a(t),\eta(t),x)\,\right] .
\eea
Thus, the $\ffi_j$ take into account the kinetic energy and $W_{a(t)}(x)$
parts of the Hamiltonian.
Next, we expand the exact solution as
$$\Psi(x,\,t,\,\hbar)\,=\,\sum_j\,b_j(\hbar,\,t)\,
\E^{iS(t)/\hbar}\,\ffi_j(A(t),\,B(t),\,\hbar,\,a(t),\,\eta(t),\,x),$$
where the $b_j(\hbar,\,t)$ satisfy an infinite linear system of ordinary
differential equations whose matrix is obtained from the
$Z_{a(t)}(x)$ term in the Hamiltonian.
In that system,
we make a first approximation by replacing the function $Z_{a(t)}(x)$ by
its Taylor approximation $Z^{[l]}_{a(t)}(x)$. This yields an infinite linear
system whose matrix is
$\widetilde{K}(l,\,t,\,\hbar)$. Its entries are time dependent polynomials
in $\hbar^{1/2}$  of order $l-1$.
We make a second approximation by truncating the infinite
system to obtain (\ref{eqcj}) that is satisfied by the
$c_j(l,\,t,\,\hbar)$. The result (\ref{est}) is proved by using
Lemma \ref{fundcalc} to show that the errors generated
by the Taylor approximation and the truncation approximation are of order
$\hbar^{l/2}$.

As described in the introduction, we construct the exponentially accurate
approximate solution
$\psi(x,\,t,\,\hbar)\,=\,\psi_{l(\hbar)}(x,\,t,\,\hbar)$, by keeping track of
the $l$--dependence of $C(l)$ in (\ref{est}) and then
choosing $l(\hbar)$ in order to minimize the error.

In the remainder of this section, we prove Theorem \ref{mainthm} under the
assumption that the following three technical lemmas are true. We prove the
first lemma in Section 3. It
estimates errors that arise from our replacement of $Z_{a(t)}(x)$ by
$Z^{[l]}_{a(t)}(x)$. The second and third lemmas are proved in Section 4. They
bound certain matrix elements and combinatorial quantities that arise from the
truncation approximation discussed above.

\vskip .25cm
\noindent
\begin{lem}\label{taylorprop}
Suppose $V$ satisfies the hypotheses of Theorem \ref{mainthm}, $|m|=l+2$,
$\widetilde{J}(l)=J+3l-3$, and
$$
\psi_l(x,\,t,\,\hbar)\,=\,\E^{iS(t)/\hbar}\
\sum_{|j|\leq\widetilde{J}(l)}\,
c_j(l,\,t,\,\hbar)\,\ffi_j(A(t),B(t),\hbar,a(t),\eta(t),x),
$$
with $\sum_{|j|\leq\widetilde{J}(l)}\,|c_j(l,\,t,\,\hbar)|^2\,=\,1$.\\
Let $\zeta(x,a(t))=(a(t)+\theta_{x,a(t)}(x-a(t)))\in \R^d$,
with $\theta_{x,a(t)}\in (0,\,1)$.
There exist constants $g_0$ and $g_1$,
that depend on $d$ and $J$ only, such that for sufficiently small $\hbar$,
\bea\nonumber
&&\left\|\,\frac{D^mV(\zeta(x,a(t)))}{m!}\ (x -a(t))^m\
\psi_{l}(\cdot,\,t,\,\hbar)\,\right\|\\ \label{goal}
&\le&g_0\,M\,\exp(4\tau(\delta^2 d+a(t)^2))\,
\left(\,g_1\,\sqrt{\hbar(l+2)}\,\| A(t)\|/\delta\,\right)^{l+2}.
\eea
\end{lem}

\vskip .25cm
\noindent
\begin{lem}\label{dlem2}
We define the infinite matrix
\bea\nonumber
&&\widetilde{X}^m_{j\,k}(t,\,\hbar)\ =
\\[5pt] \label{defx}
&&\bra\,
\ffi_j(A(t),B(t),\hbar,a(t),\eta(t),\cdot),\
(x-a(t))^m\,\ffi_k(A(t),B(t),\hbar,a(t),\eta(t),\cdot)\,\ket ,
\eea
and then define the finite matrix $X^m(l,\,t,\,\hbar)$ from
$\widetilde{X}^m(t,\,\hbar)$ by restricting its indices to
$|j|\le\widetilde{J}(l)$ and $|k|\le\widetilde{J}(l)$. Then,
$X^m_{j\,k}(l,\,s,\,\hbar)=0$ and
$\widetilde{X}^m_{j\,k}(s,\,\hbar)=0$ if $\bigg| |j|-|k|\bigg| >|m|$ and,
for each $N>0$, there exists $D(N)<\infty$, such that
\bea\nonumber
&&\sup_{\scriptsize \matrix{|k|\,\leq\widetilde{J}(l)\cr
\widetilde{J}(l)+1\le\,|j|\,\leq \widetilde{J}(l)+l+1}}
\left|\,\left(\widetilde{X}^{m_0}(t_0,\hbar)X^{m_1}(l,t_1,\hbar)
X^{m_2}(l,t_2,\hbar)\cdots X^{m_q}(l,t_q,\hbar)\right)_{j\,k}\,\right| \\
\label{dutil}&\leq&\left(\,
D(N)\,\sqrt{\hbar\,\widetilde{J}(l)}\,
\sup_{t\in\{t_0,t_1,t_2,\cdots t_q\}}\| A(t)\|\,
\right)^{|m_0|+|m_1|+|m_2|\cdots +|m_q|} ,
\eea
for any collection $m_0,\,m_1,\,\dots ,\,m_q$ of multi-indices that satisfy
$|m_j|/\widetilde{J}(l)\le N$.
\end{lem}

\vskip .25cm
\noindent
\begin{lem}\label{truncest} We define
\be\label{dcombin}
F_p(n,\,q)\,=\,\sum_{\scriptsize\matrix{1\leq |m_1|,|m_2|,\cdots,|m_q|\leq p \cr
|m_1|+|m_2|+\cdots +|m_q|=n}}1
\ee
to be the number of distinct sets $\{\,m_1,\,m_2,\,\dots,\,m_q\,\}$, where
each $m_j$ is a $d$--dimensional multi-index with $1\le |m_j|\le p$ and
$|m_1|+|m_2|+\cdots +|m_q|=n$.
We note that $F_p(n,\,q)$ is zero unless $q\le n\le qp$.
Suppose that a function $L_q(l)$ satisfies
\be\label{Lest}
L_q(l)\,\le\,\frac{C_1^q}{\hbar^q\,q!}\,
\sum_{n=3l-2}^{q(l+1)}\,(C_2\hbar l)^{n/2}\,F_{l+1}(n,\,q),
\ee
where $C_1$ and $C_2$ are constants.
Let $\grintl\,\alpha\,\grintr$ denote the greatest integer less than or equal
to $\alpha$.
Then there exists $g^*>0$, such that for any $g\in (0,\,g^*)$, there
exist positive constants $C_3$, $\gamma_1$, and $\hbar^*$, that
depend only on $g$, $C_1$, and $C_2$, such that
\be\label{lchoice}
0<\hbar<\hbar^*,\quad
l(\hbar)\,=\,\grintl\,g/\hbar\,\grintr,\quad {\mbox{and}}\quad
2\le q\le l(\hbar)+2
\ee
imply
\be
L_q(l(\hbar))\,\le\,C_3\,\exp\left\{\,-\,\gamma_1/\hbar\,\right\} .
\ee
\end{lem}

\vskip .25cm
\noindent
{\bf Proof of Theorem \ref{mainthm}}:\quad
We define $\psi_l(x,\,t,\,\hbar)$ by (\ref{app}), where the
$c_j(l,\,t,\,\hbar)$ are determined by the system (\ref{eqcj}) and initial
conditions described above.

To apply Lemma \ref{fundcalc} we define
\be\label{defxi}
\xi_l(x,\,t,\,\hbar)\,=\,
i\,\hbar\,\frac{\partial\phantom{t}}{\partial t}\psi_l(x,\,t,\,\hbar)\,-\,
\left(\,-\,\frac{\hbar^2}{2}\,\Delta\,+\,
V(x)\,\right)\,\psi_l(x,\,t,\,\hbar).
\ee
By using (\ref{profi}), we see that this can be decomposed as
a sum of two terms,
\be\label{defxi1}
\xi^{(1)}_l(x,\,t,\,\hbar)\ =\
\left(\,Z^{[l]}_{a(t)}(x)\,-\,Z_{a(t)}(x)\,\right)\,\psi_l(x,\,t,\,\hbar)
\ee
and
\be
\label{defxi2}
\xi^{(2)}_l(x,\,t,\,\hbar)\ =\
-\,P_{\{|j|\ge l+2\}}\,Z^{[l]}_{a(t)}(x)\,\psi_l(x,\,t,\,\hbar),
\ee
where $P_{\{|j|\ge l+2\}}$ is the orthogonal projection onto the span of the
set\\
$\{\,\ffi_j(A(t),\,B(t),\,\hbar,\,a(t),\,\eta(t),\,\cdot\,):\ \,|j|\ge l+2\,\}$.

By the standard Taylor series error formula,
$$Z_{a(t)}(x)\,-\,Z^{[l]}_{a(t)}(x)
\ =\ \sum_{|m|=l+2}\,\frac{D^mV(\zeta(x,a(t)))}{m!}\ (x-a(t))^m,$$
for some $\zeta(x,a)=(a+\theta_{x,a}(x-a))$, with $\theta_{x,a}\in (0,\,1)$.
Thus, by the crude estimate
\bea\nonumber
\sum_{|m|=n}\,1\ \le\ \sum_{|m|\le n}\,1&\le&
\sum_{j=1}^d\,\sum_{m_j\le n}\,1\\[5pt]
&=&(n+1)^d\ =\ \E^{d\,\ln(n+1)}\ \le\
\left(\E^d\right)^n,\label{crude}
\eea
and Lemma \ref{taylorprop}, we obtain
\bea\nonumber
&&\left\|\,\xi^{(1)}_l(\cdot,\,t,\,\hbar)\,\right\|\\[5pt]
\nonumber
&\le&\sum_{|m|=l+2}\,
\left\|\,\frac{D^mV(\zeta(x,a(t)))}{m!}\ (x-a(t))^m\
\psi_{l}(\cdot,\,t,\,\hbar)\,\right\|\\[5pt]
\nonumber
&\le&g_0\,M\,\exp(4\tau(\delta^2 d+a(t)^2))\,
\left(\,g_1\,\sqrt{\hbar(l+2)}\ \| A(t)\|/\delta\,\right)^{l+2}\,
\left(\E^d\right)^{l+2}.
\eea
Thus, there exist constants $C_4$ and $C_5$, such that $|t|\,\le\,T$ implies
\be\label{taylordone}
\left\|\,\xi^{(1)}_l(\cdot,\,t,\,\hbar)\,\right\|\,\le\,
C_4\,\left(\,C_5\,\sqrt{\hbar(l+2)}\,\right)^{l+2}.
\ee

The quantity $\xi^{(2)}_l(\cdot,\,t,\,\hbar)$ satisfies
$$
\bra\,\ffi_j(A(t),B(t),\hbar,a(t),\eta(t),x),\
\xi^{(2)}_l(x,\,t,\,\hbar)\,\ket\,=\,0,
\quad\mbox{if}\quad\left\{
\matrix{0\leq|j|\leq\widetilde{J}(l)\quad\mbox{or}\cr
|j|>\widetilde{J}(l)+l+1,}\right.
$$
and
\bea
&&\bra\,\ffi_j(A(t),B(t),\hbar,a(t),\eta(t),x),\
\xi^{(2)}_l(x,\,t,\,\hbar)\,\ket\,=
\nonumber\\[8pt]
&&\sum_{|k|\le\widetilde{J}(l)}\,
\bra\,\ffi_j(A(t),B(t),\hbar,a(t),\eta(t),x),\
Z_{a(t)}(x)\,\ffi_k(A(t),B(t),\hbar,a(t),\eta(t),x)\,\ket\,c_k(t)
\nonumber\\[6pt]
&&=(\widetilde{K}(t)\,{\bf c}(t))_j,\quad
\mbox{ if}\quad\widetilde{J}(l)<|j|\leq\widetilde{J}(l)+l+1,
\eea
where we denote the
$c_j(l,\,t,\,\hbar)$ collectively by the vector ${\bf c}(l,\,t,\,\hbar)$.
We easily verify these facts by using (\ref{eqcj}), (\ref{defk}),
(\ref{profi})
and Lemma \ref{dlem2}.
To estimate the norm of $\xi^{(2)}_l(\cdot,\,t,\,\hbar)$, we use the Dyson
expansion with remainder
to decompose
\be\label{split}
{\bf c}(l,\,t,\,\hbar)=\sum_{q=0}^l\,{\bf c}^q(l,\,t,\,\hbar)\,+\,
{\bf r}(l,\,t,\,\hbar),
\ee
where (dropping some arguments)
\be\label{cqdef}
{\bf c}^q(t)=\int_0^tds_1\int_0^{s_1}ds_2\cdots\int_0^{s_{q-1}}
ds_q(i\hbar)^{-q} K(s_1)K(s_2)\cdots K(s_q)\,
{\bf c}(0),
\ee
and
\be\label{rdef}
{\bf r}(t)=
\int_0^tds_1\int_0^{s_1}ds_2\cdots\int_0^{s_{l}}ds_{l+1}
(i\hbar)^{-(l+1)} K(s_1)K(s_2)\cdots K(s_{l+1})\,
{\bf c}(s_{l+1}).
\ee
Using (\ref{defk}), (\ref{ypoly}), and
(\ref{defx}), we see that each  ${\bf c}^q(l,\,t,\,\hbar)$ is of order
$\hbar^{q/2}$ and that ${\bf r}(l,\,t,\,\hbar)$ is of order $\hbar^{(l+1)/2}$.

To estimate the norm of $\xi^{(2)}_l(\cdot,\,t,\,\hbar)$ we study the
$j^{\mbox{\scriptsize th}}$  component of
$\widetilde{K}(t){\bf c}(l,\,t,\,\hbar)$,
with $|j|>\widetilde{J}(l)$. Because of
(\ref{split}), this coefficient is a sum of two types of terms: those that
arise from ${\bf c}^q(l,\,t,\,\hbar)$ and those that arise from
${\bf r}(l,\,t,\,\hbar)$.
Using (\ref{ypoly}), we expand $K(l,\,t,\,\hbar)$ in (\ref{cqdef}) and
(\ref{rdef}) and $\widetilde{K}(l,\,t,\,\hbar)$ in terms of
$X^m(l,\,t,\,\hbar)$ and $\widetilde{X}^m(t,\,\hbar)$, to obtain
\bea
&&\widetilde{K}(t){\bf c}^q(t) =
(i\hbar)^{-q}\sum_{n=3(q+1)}^{(l+1)(q+1)}\sum_{\scriptsize\matrix{
m_0,m_1,m_2,\cdots,m_q\cr|m_0|+|m_1|+\cdots+|m_q|=n\cr 3\leq|m_j|\leq l+1}}
\nonumber\\[9pt]
&&\int_0^tds_1\int_0^{s_1}ds_2\cdots\int_0^{s_{q-1}}ds_q
\frac{D^{m_0}V(a(t))D^{m_1}V(a(s_1))\cdots
D^{m_q}V(a(s_q))}{m_0!m_1!m_2!\cdots m_q!}\nonumber\\[4pt]
&&\qquad\qquad\qquad\qquad\qquad\times\ \widetilde{X}^{m_0}(t)
X^{m_1}(s_1)X^{m_2}(s_2)\cdots  X^{m_q}(s_q)\,{\bf c}(0),
\label{expcq}
\eea
and
\bea
&&\widetilde{K}(t){\bf r}(t) =
(i\hbar)^{-(l+1)}\sum_{n=3(l+2)}^{(l+1)(l+2)}\sum_{\scriptsize\matrix{
m_0,m_1,m_2,\cdots,m_{l+1}\cr|m_0|+|m_1|+\cdots+|m_{l+1}|=n\cr
3\leq|m_j|\leq l+1}}
\nonumber \\[9pt]
&&\int_0^tds_1\int_0^{s_1}ds_2\cdots\int_0^{s_l}ds_{l+1}
\frac{D^{m_0}V(a(t))D^{m_1}V(a(s_1))\cdots
D^{m_{l+1}}V(a(s_{l+1}))}{m_0!m_1!m_2!\cdots m_{l+1}!}\nonumber\\[4pt]
&&\qquad\qquad\qquad\qquad\qquad\times\ \widetilde{X}^{m_0}(t)
X^{m_1}(s_1)X^{m_2}(s_2)\cdots  X^{m_{l+1}}(s_{l+1})\,{\bf c}(s_{l+1}).
\label{exprt}
\eea
The values of $m_j$ that occur in both (\ref{expcq}) and
(\ref{exprt}) satisfy
\be\label{Mest}
|m_j|/\widetilde{J}(l)\le (l+1)/(3l-3)\le 1,
\ee
as long as $l\ge 3$. So, we can apply Lemma \ref{dlem2} with $N=1$.

Recall that $X^m_{j\,k}(l,\,s,\,\hbar)=0$ and
$\widetilde{X}^m_{j\,k}(s,\,\hbar)=0$ if $\bigg| |j|-|k|\bigg|>|m|$.
Since $c_k(0)$ is non-zero only for $|k|\le J$, and we need only consider
the $j^{\mbox{\scriptsize th}}$ coefficient of
$\widetilde{K}(t){\bf c}(t)$ for $|j|>\widetilde{J}(l)$, the only relevant
values of $n$ in (\ref{expcq}) must satisfy
\be\label{ndl}
n\geq 3l-2.
\ee
This condition is also satisfied for all values of $n$ in (\ref{exprt}), since
the sum begins with $n=3l+6$.

To use the analyticity assumptions to get estimates on
the derivatives of $V$, we define
$$
C_{\delta}(x)=\{z\in\C^d\,:\ z_j=x_j+\delta e^{i\theta_j},\,
\theta_j\in [0,2\pi),\,j=1,2,\cdots, d\} .
$$
If $z\in C_{\delta}(a(t))$, then, for all
$j=1,2,\cdots ,d$,
$$
|z_j|\leq\delta +|a_j(t)|.
$$
Hence, writing $\ds \frac{1}{m!}\,D^mV(a(t))$ as a $d$--dimensional Cauchy
integral, we get the bound
\be\label{mess88}
\frac{1}{m!}\,|D^mV(a(t))|\ \leq\
 \frac{\sup_{z\in C_{\delta}(a(t))} |V(z)|}{\delta^{|m|}}\ \equiv\
\frac{v(t)}{\delta^{|m|}}
\ee
where
$$
 v(t)\ \leq\ M\,\exp(4\tau(\delta^2 d+a^2(t))).
$$
Furthermore, since $\|A(t)\|$ depends continuously on $t$, there exists
$w(T)$, such that
$$\sup_{t\in [-T,T]}\ \|A(t)\|\ \leq\ w(T).$$

The norm of the vector ${\bf c}(t)$ is 1 since $K(l,\,t,\,\hbar)$ is
self-adjoint. Thus, the non-zero entries of
${\bf c}(0)$ are each bounded by 1, and by a crude estimate, there are
at most $(J+1)^d$ of them. Similarly,
${\bf c}(t)$ has at most $(\widetilde{J}(l)+1)^d$ non-zero entries, each of
which is bounded by 1.
Thus, for $l\ge 3$, (\ref{defk}), (\ref{ypoly}), (\ref{defx}),
(\ref{expcq}), (\ref{exprt}), (\ref{ndl}), (\ref{Mest}), and Lemma
\ref{dlem2} imply the following two estimates when $j$ satisfies
$\widetilde{J}(l)+1\le|j|\le\widetilde{J}(l)+l+1$:
\bea\nonumber
&&\left|\,(\widetilde{K}(t)\,{\bf c}^q(t))_{j}\,\right| \\[5pt]
&\leq&   \nonumber
v(t)\frac{\left(\int_0^Tv(s)\,ds\right)^{q}}{\hbar^q q!}\,
\sum_{n=3l-2}^{(l+1)(q+1)}\,(D(1)\,w(T)/\delta )^n\,\hbar^{n/2}\,
\widetilde{J}(l)^{n/2}\,F_{l+1}(n,q+1)\,(J+1)^d \\[6pt]\nonumber
&=&\frac{v(t)\hbar(q+1)(J+1)^d}{\int_0^Tv(s)ds}
\frac{\left(\int_0^Tv(s)ds\right)^{q+1}}{\hbar^{q+1}(q+1)!}
\sum_{n=3l-2}^{(l+1)(q+1)}(D(1)w(T)\sqrt{\hbar
\widetilde{J}(l)}/\delta)^nF_{l+1}(n,q+1)\\ \label{cqest1}
\eea
and
\bea\nonumber
&&\left|\,(\widetilde{K}(t){\bf r}(t))_{j}\,\right|\\[5pt]
\nonumber\!\! \leq&&\!\!\!\!
\frac{v(t)\hbar(l+2)(\widetilde{J}(l)+1)^d}{\int_0^Tv(s)ds}
\frac{\left(\int_0^Tv(s)ds\right)^{l+2}}{\hbar^{l+2}(l+2)!}
\sum_{n=3l-2}^{(l+1)(l+2)}(D(1)w(T)\sqrt{\hbar
\widetilde{J}(l)}/\delta )^nF_{l+1}(n,l+2),\\ \label{rest1}
\eea
where $F_p(n,q)$ is defined by (\ref{dcombin}).

By Lemma \ref{truncest}, (\ref{cqest1}), and (\ref{rest1}), both
$\left|\,(\widetilde{K}(t)\,{\bf c}^q(t))_{j}\,\right|$ and
$\left|\,(\widetilde{K}(t){\bf r}(t))_{j}\,\right|$ are bounded by
$C_3\,\exp\,\left\{\,-\gamma_1/\hbar\,\right\}$ for an appropriate choice of
$l(\hbar)\,=\,\grintl\,g/\hbar\,\grintr$ and sufficiently small $\hbar$.
Each of the $l+2$ terms in (\ref{split}) contributes a term of this type to
$\xi^{(2)}_l(x,\,t,\,\hbar)$, so
\bea\nonumber
\hbar^{-1}\ \left\|\,\xi^{(2)}_l(x,\,t,\,\hbar)\,\right\|
&\le&C_3\,\hbar^{-1}\,(l(\hbar)+2)\,\exp\,\left\{\,-\,\gamma_1/\hbar\,\right\}
\\[5pt] \nonumber
&\le&C_3\,\exp\,\left\{\,-\,\gamma_2/\hbar\,\right\} ,
\eea
for any $\gamma_2<\gamma_1$ when $\hbar$ is sufficiently small.

We shrink $g$ if necessary to make
\be\label{george}
C_5^2\,g<1
\ee
and set $l=l(\hbar)$ in
(\ref{taylordone}). This yields a similar estimate
$$\hbar^{-1}\ \left\|\,\xi^{(1)}_l(\cdot,\,t,\,\hbar)\,\right\|\,\le\,
C_6\,\exp\,\left\{\,-\,\gamma_3/\hbar\,\right\} ,$$
for some $\gamma_3>0$.

We combine these two estimates and apply Lemma \ref{fundcalc} to obtain
(\ref{expest}) with\\ $\gamma\,=\,\min\,\{\gamma_2,\,\gamma_3\}$.
This proves the theorem. \ep

\vskip .5cm
\section{Proof of Lemma 2.2}
\setcounter{equation}{0}

For simplicity, we drop the $t$ dependence in the notation throughout this
section.

To prove Lemma \ref{taylorprop}, we use H\"older's inequality to see that
$\sum_{|j|\leq\widetilde{J}(l)}\,|c_j(l,\,\hbar)|^2\,=\,1$
implies
\bea\nonumber
\sum_{|j|\leq\widetilde{J}(l)}\,|c_j(l,\,\hbar)|&\le&
\left(\,\sum_{|j|\leq\widetilde{J}(l)}\,|c_j(l,\,\hbar)|^2\,\right)^{1/2}\,
\left(\,\sum_{|j|\leq\widetilde{J}(l)}\,1\,\right)^{1/2}\\ \nonumber
&\le&\left(\,\widetilde{J}(l)+1\,\right)^{d/2}.
\eea
Thus, it is sufficient to prove
\bea\nonumber
&&\left\|\,\frac{D^mV(\zeta(x,a))}{m!}\ (x-a)^m\
\ffi_{j}(A,\,B,\,\hbar,\,a,\,\eta,\,x)\,\right\|
\\ \label{goal1}
&\le&g_3\,M\,\exp(4\tau(\delta^2 d+a^2))\,
\left(\,g_4\,\sqrt{\hbar(l+2)}\,\| A\|/\delta\,\right)^{l+2},
\eea
for some $g_3$ and $g_4$, whenever $|j|\leq\widetilde{J}(l)$.

We mimic the proof of \ref{mess88} to obtain a bound on
$D^mV(\zeta(x,a))/m!$.
If $z\in C_{\delta}(\zeta(x,a))$, then, for all
$j=1,2,\cdots ,d$,
$$
|z_j|\,\leq\,\delta +|\zeta_j(x,a)|\,\leq\,\delta +|a_j|+|x_j-a_j|.
$$
Using this and applying $(b+c)^2\leq 2(b^2+c^2)$ several times,
we see that
$z\in C_{\delta}(\zeta(x,a))$ implies
$$
|\,V(z)\,|\,\leq\,M\,\exp(2\tau(x-a)^2)\,\exp(4\tau(\delta^2 d+a^2)).
$$
Hence, writing $\ds \frac{1}{m!}\,D^mV(\zeta(x,a))$ as a $d$--dimensional Cauchy
integral, we easily obtain the bound
\be\label{mess8}
\frac{1}{m!}\,|D^mV(\zeta(x,a))|\leq
M\ \frac{\exp(4\tau(\delta^2 d+a^2))}{\delta^{|m|}}\
\exp(2\tau(x-a)^2).
\ee
Thus, estimate (\ref{goal1}) follows from the corresponding estimate
on the integral
\be\nonumber
I=\int_{\R^d}\,(x-a)^{2m}\,\exp(4\tau(x-a)^2)\
|\ffi_{j}(A,\,B,\,\hbar,\,a,\,\eta,\,x)|^2\,dx.
\ee
Performing the change of variables $x\mapsto y=|A|^{-1}(x-a)/\hbar^{1/2}$,
and using the explicit formula for $\ffi_j$, we see that
\be\nonumber
I\,=\,\frac{\hbar^{|m|}}{2^{|j|}\,j!\,\pi^{d/2}}\,\int_{\R^d}\,
(|A|y)^{2m}\,\exp(-y^2+4\tau\hbar(|A|y)^2)\ |{\cal H}_j(A;\,y)|^2\,dy,
\ee
where ${\cal H}_l(A;\,y)$ is the polynomial described in Remarks 4 and 5
that immediately follow the definition of $\ffi_{j}(A,B,\hbar,a,\eta,x)$ in
Section 2.

We assume henceforth that $\hbar$ is sufficiently small that
\be\label{35}
4\,\tau\,\hbar\,\| A\|^2\,\leq\,1/2.
\ee
The estimate $(|A|y)^2\,\leq\,\| A\|^2\,y^2$
implies $(|A|y)_k^2\,\leq\,\| A\|^2\,y^2$,
for $k=1,2,\cdots, d$. From this, we conclude
\be\label{Iest}
I\ \leq\ \frac{{(\hbar\,\| A\|^2)}^{|m|}}{2^{|j|}\,j!\,\pi^{d/2}}\ \int\,
y^{2|m|}\,\exp(-y^2/2)\,|{\cal H}_j(A;\,y)|^2\,dy.
\ee
We next need an estimate on  $|{\cal H}_j(A;\,y)|$.
By Remark 5 after the definition in Section 2,
\be\label{amazing}
\left(-\Delta+y^2\right)\,{\cal H}_j(A;\,y)\,\E^{-y^2/2}
=(2|j|+d)\ {\cal H}_j(A;\,y)\,\E^{-y^2/2}.
\ee
This equation states that ${\cal H}_j(A;\,y)\,\E^{-y^2/2}$ is a
eigenfunction of $-\Delta+y^2$ corresponding
to the eigenvalue $2|j|+d$. Introducing normalization factors, we conclude
that
\be\nonumber
2^{-|j|/2}\,(j!)^{-1/2}\,\pi^{-d/4}\,
{\cal H}_j(A;\,y)\,\E^{-y^2/2}\,=\,\sum_{|k|=|j|}\,
b_k\,\ffi_{k_1}(y_1)\ffi_{k_2}(y_2)\cdots \ffi_{k_d}(y_d),
\ee
where $\ffi_{k}(y)\,=\,2^{-k/2}\,(k!)^{-1/2}\,\pi^{-1/4}\,H_k(y)\,\E^{-y^2/2}$
is the normalized eigenfunction of
$\ds -\frac{\partial^2\phantom{y}}{\partial y^2}+y^2$
corresponding to the eigenvalue $2k+1$, and the coefficients $b_k$ satisfy
$\sum_{|k|=|j|}\,|b_k|^2\,=\,1$.
We can thus deduce an estimate of ${\cal H}_j(A;\,y)$ from an estimate for
usual Hermite polynomials $ H_k(y)$:
\bea\nonumber
|{\cal H}_j(A;\,y)|^2&=&|\sum_{|k|=|j|}\,
b_k\,H_{k_1}(y_1)H_{k_2}(y_2)\cdots H_{k_d}(y_d)|^2\\[5pt]
&\leq&\sum_{|k|=|j|}\,|H_{k_1}(y_1)H_{k_2}(y_2) \cdots H_{k_d}(y_d)|^2.
\label{mess18}
\eea
The Hermite polynomials, in turn, satisfy the following bounds.

\vskip .25cm
\noindent
\begin{lem}\label{taylor1}
If $|y|>\sqrt{2k+1}$, then
\be\label{joyest}
|H_k(y)|\,\leq\,2^k\,|y|^k,
\ee
and there exists a numerical constant $\kappa >0$, such that for all $y\in\R$,
\be\nonumber
|H_k(y)|\,\leq\,\kappa\,2^{k/2}\,\sqrt{k!}\,\E^{y^2/2}.
\ee
\end{lem}

\vskip .25cm
\noindent
{\bf Proof:}\quad
The second statement is well known. See \cite{AandS}, formula 22.14.17 or
\cite{GrandRy5}, formula 8.954.2.

By symmetry, it is sufficient to prove (\ref{joyest}) for $y>0$.
It is well known that the $k^{\mbox{\scriptsize th}}$ eigenfuction of the
harmonic oscillator is non-zero in the classically forbidden
region $|y|>\sqrt{2k+1}$.
It is also well known that $H_k(y)=2^ky^k+P_k(y)$
where $P_k(y)$ is a polynomial of degree $k-2$. Thus, we have
$$
H_k(y)/(2^ky^k)=1+B_k(y),
$$
where $B_k$ satisfies
$$
B_k(y)\,=\,O(1/y^2)\qquad \mbox{for large } y,
$$
and
$$
B_k(y)>-1\qquad \mbox{if}\quad y>\sqrt{2k+1}.
$$
Using standard properties of the Hermite polynomials, we see that
\bea\nonumber
B_k'(y)&=&(y\,H_k'(y)\,-\,k\,H_k(y))/(2^k\,y^{k+1})\\[5pt] \nonumber
&=&(2\,k\,y\,H_{k-1}(y)\,-\,k\,H_k(y))/(2^k\,y^{k+1})\\[5pt] \nonumber
&=&k\,(2\,y\,H_{k-1}(y)\,-\,(2\,y\,H_{k-1}(y)\,-\,
2\,(k-1)\,H_{k-2}(y)))/(2^k\,y^{k+1})\\[5pt] \nonumber
&=&k\,(k-1)\,H_{k-2}(y)/(2^{k-1}\,y^{k+1}).
\eea
So, if $y>\sqrt{2k+1}$, we see that $B_k'(y)>0$.
Thus, $y>\sqrt{2k+1}$ implies $-1<B_k(y)<0$,
which implies (\ref{joyest}). \ep

\vskip .25cm
In (\ref{Iest}), we use the multinomial expansion
$$
(y^2)^{|m|}\,=\,\sum_{|n|=|m|}\,\pmatrix{&|m|&\cr n_1&\cdots&n_d}
\,y_1^{2n_1}\,y_2^{2n_2}\,\cdots\,y_d^{2n_d}
$$
to obtain
$$
I\leq\frac{{(\hbar\,\| A\|^2)}^{|m|}}{2^{|j|}\,j!\,\pi^{d/2}}
\sum_{|n|=|m|}\sum_{|k|=|j|}\pmatrix{&|m|& \cr n_1&\cdots&n_d}\prod_{r=1}^d
\int y_r^{2n_r}|H_{k_r}(y_r)|^2\exp(-y_r^2/2) dy_r.
$$
Using Lemma \ref{taylor1},
the one dimensional integrals in the final factor here can then be estimated as
follows (without indices):
\bea\nonumber
&&\int\,y^{2n}\,|H_{k}(y)|^2\,\exp(-y^2/2)\,dy\\[7pt]
&\le&\int_{y^2\leq 2k+1}\kappa^2\,2^k\,k!\,y^{2n}\,\exp(y^2/2)\,dy\,+\,
  \int\,4^k\,y^{2(n+k)}\,\exp(-y^2/2)\,dy\nonumber\\[7pt]
&\le&\kappa^2 2^k k! \frac{2}{2n+1} (2k+1)^{n+1/2} \exp(k+1/2)\,+\,
  4^k 2^{(k+n+1/2)} \int z^{2(k+n)}e^{-z^2}dz
\nonumber\\[7pt]
&=&\frac{2^{k+1}}{2n+1} \kappa^2 e^{k+1/2} k! (2k+1)^{n+1/2}\,+\,
  4^k 2^{(k+n+1/2)} \pi^{1/2} \frac{(2(k+n)-1)!}{2^{2(n+k)-1}(k+n-1)!}
\nonumber\\[7pt]
&=&\frac{2^{k+1}}{2n+1} \kappa^2 e^{k+1/2} k! (2k+1)^{n+1/2}\,+\,
  2^{k-n+3/2} \pi^{1/2} \frac{(2(k+n)-1)!}{(k+n-1)!}.\label{mess24}
\eea
Here we have used
$$
\int\,z^{2(k+n)}\,e^{-z^2}\,dz\ =\ \pi^{1/2}\,
\frac{1\cdot 3\cdot 5\cdots (2(k+n)-1)}{2^{k+n}}.
$$
Now, Stirling's formula guarantees the
existence of $a>0$, such that
\be\label{stirling}
a^n\,n^n\leq n!\leq n^n,
\ee
for all integers $n\ge 1$.
Using this in (\ref{mess24}), we obtain
$$
\int\,y^{2n}\,|H_{k}(y)|^2\,\exp(-y^2/2)\,dy\le f_0\,f_1^{k+n}\,(k+n)!,
$$
for some constants $f_0$ and $f_1$.
From this we conclude (restoring the indices)
\be\label{mess26}
I\ \le\ \frac{{(\hbar\,\| A\|^2)}^{|m|}}{2^{|j|}\,j!\,\pi^{d/2}}\,
\sum_{|n|=|m|}\ \sum_{|k|=|j|}\,f_0^d\,f_1^{|m|+|j|}\
\pmatrix{&|m|& \cr n_1&\cdots&n_d}\,(k+n)!
\ee
In this expression, we have
\bea\nonumber
(k+n)!&=&(|k|+|n|)!\,
\pmatrix{&|k|+|n|&\cr k_1+n_1&\cdots&k_d+n_d}^{-1}\\[5pt] \nonumber
&=&(|m|+|j|)!\,
\pmatrix{&|m|+|j|&\cr k_1+n_1&\cdots&k_d+n_d}^{-1}\\[7pt]
&\le&(|m|+|j|)!
\eea
We further use
$\ds \sum_{|n|=|m|}\,\pmatrix{&|m|& \cr n_1 & \cdots & n_d}=d^{|m|}$
and (\ref{crude}) in (\ref{mess26}) to obtain
\bea
&&I\ \leq\ (\hbar\,\| A\|^2)^{|m|}\,
\left(\,\frac{f_0}{\sqrt{\pi}}\,\right)^d\,
(f_1\,d)^{|m|}\,
\left(\,\frac{\E^d\,f_1}{2}\right)^{|j|}\,
\frac{(|m|+|j|)!}{j!}\nonumber\\[7pt]
&&=\ (\hbar\,\| A\|^2)^{|m|}\,
\left(\,\frac{f_0}{\sqrt{\pi}}\,\right)^d\,
(f_1\,d)^{|m|}\,
\left(\,\frac{\E^d\,f_1}{2}\right)^{|j|}\,
\pmatrix{&|j|&\cr j_1&\cdots &j_d}\,
\pmatrix{|m|+|j|\cr |j|}\,|m|!\nonumber\\[7pt]
&&\le\ (\hbar\,\| A\|^2)^{|m|}\,
\left(\,\frac{f_0}{\sqrt{\pi}}\,\right)^d\,
(2\,f_1\,d)^{|m|}\,
(\E^d\,f_1)^{|j|}\,
\pmatrix{&|j|&\cr j_1&\cdots &j_d}\,|m|!\nonumber\\[7pt]
&&\equiv(\hbar\,\| A\|^2 d)^{|m|}\,f_2^d\,f_3^{|m|}\,f_4^{|j|}\,
\pmatrix{&|j|&\cr j_1&\cdots &j_d}\,|m|!,\label{mess30}
\eea
where $f_2$, $f_3$, $f_4$ are numerical constants.

Estimates (\ref{mess8}) and (\ref{mess30}) imply
\bea\nonumber
&&\left\|\,\frac{D^mV(\zeta(x,a))}{m!}\,(x-a)^m\,
\ffi_{j}(A,\,B,\,\hbar,\,a,\,\eta,\,x)\,\right\|
\\ \nonumber
&&\le\,M\exp(4\tau(\delta^2 d+a^2))
(\hbar\| A\|^2(d/\delta^2)f_3)^{|m|/2}
f_2^{d/2}f_4^{|j|/2}\sqrt{|m|!}
\sqrt{\pmatrix{&|j|&\cr j_1&\cdots &j_d}}.
\eea
So, by the Schwarz inequality, (\ref{app}), (\ref{mess18}), and (\ref{silly}),
\bea
&&\left\|\,\frac{D^mV(\zeta(x,a))}{m!}\,(x-a)^m\,\psi_{l}(x,t, \hbar)\,\right\|^2
\nonumber\\[5pt]
&&\le\,\sum_{n=0}^{\widetilde{J}(l)}\,\sum_{|j|=n}\,
\left\|\,\frac{D^mV(\zeta(x,a))}{m!}\,(x-a)^m\,
\ffi_{j}(A,\,B,\,\hbar,\,a,\,\eta,\,x)\,\right\|^2\nonumber\\[5pt]
&&\le\,M^2\exp(8\tau(\delta^2 d+a^2))
(\hbar\| A\|^2(d/\delta^2)f_3)^{|m|}f_2^{d}|m|!\,
\sum_{n=0}^{\widetilde{J}(l)}
\sum_{|j|=n}f_4^{|j|}\pmatrix{&|j|&\cr j_1&\cdots &j_d}\nonumber\\[5pt]
&&=\,M^2\,\exp(8\tau(\delta^2 d+a^2))\,
(\hbar\| A\|^2(d/\delta^2)f_3)^{|m|}\,f_2^{d}\,|m|!\,
\sum_{n=0}^{\widetilde{J}(l)}\,f_4^{n}\,d^n\nonumber\\[5pt]
&&=\,M^2\,\exp(8\tau(\delta^2 d+a^2))\,
(\hbar\| A\|^2(d/\delta^2)f_3)^{|m|}\,f_2^{d}\,|m|!\,
\left(\,\frac{(f_4d)^{\widetilde{J}(l)+1}-1}{f_4d-1}\,\right)\nonumber\\[5pt]
&&\le\,M^2\,\exp(8\tau(\delta^2 d+a^2))\,
(\hbar\| A\|^2(d/\delta^2)f_3)^{|m|}\,f_2^{d}\,|m|!\,
2\,(f_4 d)^{\widetilde{J}(l)}.
\eea

The last step depends on the following fact: We can assume without
loss that $q=f_4d\ge 2$, so that for any positive integer $p$, we have
$-1\le q^{p+1}-2q^p$, and hence, $q^{p+1}-1\le 2q^{p+1}-2q^p$.
Thus,
\be\label{silly}
(q^{p+1}-1)/(q-1)\le 2q^p.
\ee

We use the hypothesis $|m|=l+2$ and note that $|m|!\,\le\,(l+2)^{l+2}$.
We then conclude that there exist constants $g_0$ and $g_1$,
that depend only on $d$ and $J$, such that (\ref{goal}) holds. This implies the
lemma. \ep

\vskip .5cm
\section{Proofs of Lemmas 2.3 and 2.4}
\setcounter{equation}{0}

The proof of Lemma 2.3 relies on two preliminary lemmas.

\vskip .25cm
\noindent
\begin{lem}\label{dlem0}
The matrix elements of $(x-a)^m$ satisfy
\bea\nonumber
&&\left|\,\langle\,\ffi_j(A,\,B,\,\hbar,\,a,\,\eta,\,x),\,(x-a)^m\,
\ffi_k(A,\,B,\,\hbar,\,a,\,\eta,\,x)\rangle\,\right|\\[7pt]
&\leq&\hbar^{|m|/2}\,(\sqrt{2}d)^{|m|}\,\| A\|^{|m|}\,
\sqrt{(|k|+1)(|k|+2)\cdots (|k|+|m|)},
\eea
and
$$
\langle\,\ffi_j(A,\,B,\,\hbar,\,a,\,\eta,\,x),\,(x-a)^m\,
\ffi_k(A,\,B,\,\hbar,\,a,\,\eta,\,x)\,\rangle\,=\,0,
\quad\mbox{if}\quad\bigg| |j|-|k|\bigg| >|m|.
$$
\end{lem}

\vskip .25cm
\noindent
{\bf Proof:}\quad
For $i=1,2,\cdots, d$, we can use equation (3.28) of \cite{raise} to express
$(x_i-a_i)$ in terms of raising and lowering operators. Doing so, we obtain
\bea\nonumber
(x_i-a_i)\,\ffi_k(A,\,B,\,\hbar,\,a,\,\eta,\,x)
&=&\sqrt{\hbar/2}\,\sum_{p=1}^d\,A_{ip}\,\sqrt{k_p+1}\,
\ffi_{k'(p)}(A,\,B,\,\hbar,\,a,\,\eta,\,x)\\
\nonumber &&+\,\sqrt{\hbar/2}\,\sum_{p=1}^d\,\bar{A}_{ip}\,\sqrt{k_p}\,
\ffi_{k''(p)}(A,\,B,\,\hbar,\,a,\,\eta,\,x),
\eea
where $k'(p)$ and $k''(p)$ are constructed from $k$ by replacing
the component $k_p$ by $k_p+1$ and $k_p-1$, respectively.
Thus, $(x_i-a_i)\,\ffi_k(A,\,B,\,\hbar,\,a,\,\eta,\,x)$ may be written as a sum
of $2d$ terms, each of which has the form
$b_q\,\ffi_q(A,\,B,\,\hbar,\,a,\,\eta,\,x)$, where $|q|\,\le\,|k|+1$ and
$|b_q|\,\le\,\sqrt{\hbar/2}\,\|A\|\,\sqrt{|k|+1}$.

From this and a simple induction,
$(x-a)^m\,\ffi_k(A,\,B,\,\hbar,\,a,\,\eta,\,x)$
may be written as a sum of $(2d)^{|m|}$ terms, each of which has the form
$b_q\,\ffi_q(A,\,B,\,\hbar,\,a,\,\eta,\,x)$, where
$|k|-|m|\,\le\,|q|\,\le\,|k|+|m|$ and
$|b_q|\,\le\,(\hbar/2)^{|m|}\,\|A\|^{|m|}\,
\sqrt{(|k|+1)(|k|+2)\cdots(|k|+|m|)}.$

This implies the lemma. \ep

\vskip .25cm
In the next lemma we use the shorthand $X^m(t)$ to denote $X^m(l,\,t,\,\hbar)$ of
Lemma \ref{dlem2}.

\vskip .25cm \noindent
\begin{lem}\label{dlem1}
For each $N>0$, there exists $D(N)<\infty$, such that
\bea\nonumber
&&\sup_{\{j,k\,:\ |j|,\,|k|\leq \widetilde{J}(l)\}}
\left|\,(X^{m_1}(t_1)\,X^{m_2}(t_2)\,\cdots\,X^{m_q}(t_q))_{j\,k}\,\right|
\\[5pt]\label{muly}&\leq&\left(D(N)\,\sqrt{\hbar}
\sup_{t\in\{t_1,t_2,\cdots t_q\}}\| A(t)\|\right)^{|m_1|+|m_2|\cdots +|m_q|}\,
\widetilde{J}(l)^{(|m_1|+|m_2|+\cdots +|m_q|)/2},
\eea
for any collection $m_1,\,m_2,\,\dots ,\,m_q$ of multi-indices that satisfy
$|m_j|/\widetilde{J}(l)\le N$.
\end{lem}

\vskip .25cm \noindent
{\bf Proof:}\quad
The matrix $X^m$ is obtained from $\widetilde{X}^m$ by restricting its indices
to $|j|,\,|k|\le\widetilde{J}(l)$.
Using Lemma \ref{dlem0}, we see that for such values of the indices,
$$
\left|\,X^m_{j\,k}(t)\,\right|\le
(\sqrt{2\hbar}\,d\,\| A(t)\|)^{|m|}\,\widetilde{J}(l)^{|m|/2}\,
\sqrt{(1+1/\widetilde{J}(l))(1+2/\widetilde{J}(l)))\cdots
(1+|m|/\widetilde{J}(l))}.
$$
Thus, $|m|/\widetilde{J}(l)\le N$ implies
$$
\left|\,X^m_{j\,k}(t)\,\right|\,
\le\,(\sqrt{2\hbar}\,d\,\| A(t)\|)^{|m|}\,(1+N)^{|m|/2}\,
\widetilde{J}(l)^{|m|/2}.
$$

We estimate the absolute value of the $j,\,k$ matrix element of the product
\bea\nonumber
&&\left|\,(X^{m_1}(t_1)\,X^{m_2}(t_2)\,\cdots\,X^{m_q}(t_q))_{j\,k}\,\right|
\\[10pt]\nonumber
&\le&\sum_{\scriptsize
\matrix{|p_i|\leq \widetilde{J}(l),\; i=1,\cdots , q-1\cr
|p_1-j|\leq |m_1|, |p_2-p_1|\leq |m_2|, \cdots, |k-p_{q-1}|\leq |m_q|}}
\left|\,X^{m_1}_{j\,p_1}(t_1)\,X^{m_2}_{p_1\,p_2}(t_2)\,\cdots\,
X^{m_q}_{p_{q-1}\,k}(t_q)
\,\right|\\ \label{dprod1}
\eea
by bounding the absolute values of each of the matrix elements
$X^{m_i}_{p_{i-1}\,p_i}(t_i)$ by
\be\label{dprod2}
\left|\,X^{m_i}_{p_{i-1}\,p_i}(t_i)\,\right|\,\leq
(\sqrt{2\hbar}\,d\,\| A(t_i)\|)^{|m_i|}\,(1+N)^{|m_i|/2}\,
\widetilde{J}(l)^{|m_i|/2},
\ee
and multiplying by a bound on the number of terms.

To estimate the number of terms, we first note that for any multi-index $r$,
the number of multi-indices indices $p$ that satisfy
$\bigg| |p|-|r|\bigg| \leq |m|$ is equal to the number of vectors
$v$ with integer components, such that $|\sum_{i=1}^d\,v_i|\leq |m|$.
This is bounded by the number of vectors with integer components, such
that $|v_i|\leq |m|$ for
$i=1,2,\cdots, d$, which is $(2|m|+1)^d$.

Thus, the number of terms is bounded by
\bea\nonumber
&&\sum_{\scriptsize
\matrix{|p_i|\leq\widetilde{J}(l),\, i=1,\cdots , q-1\cr
|p_1-j|\leq |m_1|, |p_2-p_1|\leq |m_2|,
\cdots, |k-p_{q-1}|\leq |m_q|}}1
\\[10pt]\nonumber
&\leq&(2|m_1|+1)^d(2|m_2|+1)^d\cdots(2|m_q|+1)^d
\\[10pt]\label{dprod3}
&\leq&(e^2)^{d(|m_1|+|m_2|+\cdots +|m_q|)}.
\eea
The final inequality follows because $1+2|m_j|\le e^{2|m_j|}$.

We obtain the lemma by using (\ref{dprod2}) and (\ref{dprod3}) to bound
(\ref{dprod1}).
\ep

\vskip .25cm \noindent
{\bf Proof of Lemma \ref{dlem2}}\quad
We mimic the proof of Lemma \ref{dlem1}.
Let $\widetilde{m}=|m_0|+|m_1|+|m_2|+\dots+|m_q|$.
By Lemmas \ref{dlem0} and \ref{dlem1},
\bea
&&\left(\sqrt{\hbar}
\sup_{t\in\{t_0,t_1,\cdots, t_q\}}\| A(t)\|\right)^{-\widetilde{m}}\
\left|\,
(\widetilde{X}^{m_0}(t_0)\,X^{m_1}(t_1)\,X^{m_2}(t_2)\,
\cdots\,X^{m_q}(t_q))_{j\,k}\,\right|
\nonumber\\[12pt]
&=&\left(\sqrt{\hbar}
\sup_{t\in\{t_0,t_1,\cdots, t_q\}}\| A(t)\|\right)^{-\widetilde{m}}
\sum_{|r|\le\widetilde{J}(l)}
\left|\,
\widetilde{X}^{m_0}_{j\,r}(t_0)\,\right|\,\left|\,
(X^{m_1}(t_1)\,X^{m_2}(t_2)\,\cdots\,X^{m_q}(t_q))_{r\,k}\,\right|
\nonumber\\[12pt]
&\le&\sum_{\scriptsize \matrix{r\le\widetilde{J}(l)\cr |j-r|\le|m_0|}}
2^{|m_0|/2}\,\sqrt{(|j|+1)(|j|+2)\cdots (|j|+|m_0|)}
\nonumber\\&&\qquad\qquad\qquad\qquad\qquad\qquad\times\,
D(N)^{|m_1|+|m_2|+\dots +|m_q|}\,
\widetilde{J}(l)^{(|m_1|+|m_2|+\dots +|m_q|)/2}
\nonumber\\[15pt]
&\le&\sum_{\scriptsize \matrix{|r|\widetilde{J}(l)\cr |j-r|\le|m_0|}}
2^{|m_0|/2}\,\widetilde{J}(l)^{|m_0|/2}\,
\sqrt{
 (1+1/\widetilde{J}(l))(1+2/\widetilde{J}(l))\dots(1+|m_0|/\widetilde{J}(l))}
\nonumber\\&&\qquad\qquad\qquad\qquad\qquad\qquad\times\,
D(N)^{|m_1|+|m_2|+\dots +|m_q|}\,
\widetilde{J}(l)^{(|m_1|+|m_2|+\dots +|m_q|)/2}
\nonumber\\[15pt]
&\le&\sum_{\scriptsize \matrix{|r|\le\widetilde{J}(l)\cr |j-r|\le |m_0|}}
2^{|m_0|/2}\,(1+N)^{|m_0|/2}\,
D(N)^{|m_1|+|m_2|+\dots +|m_q|}\,
\widetilde{J}(l)^{(|m_1|+|m_2|+\dots +|m_q|)/2}
\nonumber\\[12pt]
&\le&2^{|m_0|/2}\,(1+N)^{|m_0|/2}\,
D(N)^{|m_1|+|m_2|+\dots +|m_q|}\,
\widetilde{J}(l)^{(|m_1|+|m_2|+\dots +|m_q|)/2}\,
\sum_{|j-r|\le |m_0|}\, 1.\nonumber
\eea
Since the sum in the last line is $(1+2|m_0|)^d\,\le\,e^{2d|m_0|}$, we
obtain the desired estimate.
\ep

\vskip .25cm
We now turn to the proof of Lemma \ref{truncest}. Our first step is to study the
combinatorial factor $F_p(n,\,q)$ in the one dimensional case.

\vskip .25cm
\noindent
\begin{lem}\label{lem3}
Let $m_j$ denote positive integers, and let
\be\label{combi}
G_p(n,\,q)=\sum_{\scriptsize\matrix{1\leq m_1,m_2,\cdots,m_q\leq p \cr
m_1+m_2+\cdots +m_q=n}}1.
\ee
This quantity is zero if $n<q$ or $n>qp$. Otherwise, it satisfies
\be\label{combin}
G_p(n,\,q)\leq \left\{
\matrix{\pmatrix{n-1\cr q-1}&\mbox{if}&q\leq n\leq [q(p+1)/2]\cr
        \pmatrix{q(p+1)-n-1\cr q-1}&\mbox{if}&[q(p+1)/2]+1\leq n\leq qp}\right.
\ee
\end{lem}

\vskip .25cm
\noindent
{\bf Proof:}\quad
If $n<q$ or $n>qp$, the sum in (\ref{combi}) contains no terms, so
$G_p(n,\,q)=0$.

The quantity $G_p(n,\,q)$ is the number of ways that the number $n$ can be
decomposed as $n=m_1+m_2+\dots +m_q$ with each $m_j$ satisfying $1\le m_j\le p$.
We uniquely associate to each such decomposition of $n$, a corresponding
decomposition of $n'=q(p+1)-n=m_1'+m_2'+\dots +m_q'$ by setting
$m_j'=p+1-m_j$. This association is a one-to-one correspondence, so we see that
$G_p(n,\,q)$ satisfies the symmetry relation
\be\label{gsym}
G_p(n,\,q)=G_p(q(p+1)-n,\,q).
\ee
As a consequence, the second inequality in (\ref{combin})
(with $[q(p+1)/2]+1\leq n\leq qp$) follows from the first
(with $q\le n\le [q(p+1)/2]$).

To prove the first inequality in (\ref{combin}) we drop the condition
$m_j\le p$ and exactly calculate the resulting function.
Dropping the upper bound on the $m_j$, we clearly have
$G_p(n,\,q)\le G(n,\,q)$, where
$$
G(n,\,q)=\sum_{\scriptsize\matrix{1\leq m_1,m_2,\cdots,m_q\cr
m_1+m_2+\cdots +m_q=n}}1.
$$
So, the lemma will be proved once we establish
\be\label{gexact}
G(n,\,q)=\pmatrix{n-1\cr q-1}.
\ee

To prove this, we use induction.
Formula (\ref{gexact}) is trivial to verify for all $n\ge 1$ when $q=1$ and for
all $n\ge 2$ when $q=2$.
Assume (\ref{gexact}) has been verified whenever $n\ge q-1$, and suppose
$n\ge q$.
We have
$$
G(n,\,q)\,
=\,\sum_{m_1=1}^{n-q+1}\,G(n-m_1,\,q-1)
\,=\,\sum_{m_1=1}^{n-q+1}\,\pmatrix{n-m_1-1\cr q-2},
$$
so the induction step will be complete once we show
\be\label{mystar}
\sum_{m_1=1}^{n-q+1}\,\pmatrix{n-m_1-1\cr q-2}
\,=\,\pmatrix{n-1\cr q-1}.
\ee

To prove (\ref{mystar}), we do an another induction on $n\ge q$.
For $n=q$ we trivially have
$$
\sum_{m=1}^1\,\pmatrix{q-2\cr q-2}\,=\,1\,=\,\pmatrix{q-1\cr q-1}.
$$
Now assume (\ref{mystar}) is true for some $n\ge q$. For the case of $n+1$, we
have

\bea
\nonumber
&&\sum_{m=1}^{(n+1)-q+1}\,\pmatrix{(n+1)-m-1\cr q-2}
\\[7pt] \nonumber
&=&\sum_{m=1}^{n-q+1}\,\pmatrix{n-m\cr q-2}
\quad +\quad\pmatrix{(n+1)-\{(n+1)-q+1\}-1\cr q-2}
\\[7pt] \nonumber
&=&\sum_{m=1}^{n-q+1}\,\pmatrix{n-m\cr q-2}
\quad +\quad\pmatrix{q-2\cr q-2}
\\[7pt] \nonumber
&=&\sum_{m=1}^{n-q+1}\,\pmatrix{n-m\cr q-2}
\quad +\quad 1
\\[7pt] \nonumber
&=&\sum_{m'=0}^{n-q}\,\pmatrix{n-m'-1\cr q-2}
\quad +\quad 1
\\[7pt] \nonumber
&=&\pmatrix{n-1\cr q-2}\ +\
\sum_{m'=1}^{n-q+1}\,\pmatrix{n-m'-1\cr q-2}\ -\
\pmatrix{q-2\cr q-2}\ +\ 1
\\[7pt] \nonumber
&=&\pmatrix{n-1\cr q-2}\ +\ \pmatrix{n-1\cr q-1}
\\[7pt] \nonumber
&=&\pmatrix{n\cr q-1}.
\eea
This proves (\ref{mystar}) and completes the proof of the lemma.
\ep

\vskip .35cm
Our next step is to obtain an estimate for $G_p(n,\,q)$ that depends only on
$n$.

\vskip .25cm
\noindent
\begin{lem}\label{lem4}
Let $G_p(n,\,q)$ be defined by (\ref{combi}).
There exists a constant $C_7$, such that
$p\ge 1$, $q\ge 1$, and $n\ge 1$ imply
\be\label{escom}
G_p(n,q))\leq (C_7)^n.
\ee
\end{lem}

\vskip .25cm
\noindent
{\bf Proof:}\quad
We note that $G_p(n,q)$ is zero unless $q\le n\le qp$.
Furthermore, $G_p(q,q)=G_p(qp,q)=G_2(2,1)=1$ and $G_1(n,q)=\delta_{n\,q}$.
So, we need only prove existence of $C_7$,
such that $p\ge 2$, $q\ge 1$, $(p,q)\ne (2,1)$ imply
\be\label{escom1}
\sup_{q<n<pq}\ (G_p(n,q))^{1/n}\,\leq\,C_7.
\ee
To prove this, we first study the case where $q<n\leq [q(p+1)/2]$.
Since $q\leq n$, we have
$$
q^{1/n}\,=\,\exp(\ln(q)/n)\,\le\,\exp(\ln(n)/n)\,\le\,\E^{1/\e}.
$$
Thus, by (\ref{combin}), the condition $q<n\leq [q(p+1)/2]$,
and (\ref{stirling}), we have
\bea\nonumber
(G_{p}(n,\,q))^{1/n}&\le&\left(\,\frac{(n-1)!}{(q-1)!\,(n-q)!}\,\right)^{1/n}\\
\nonumber&\le& q^{1/n}\ \left(\,\frac{(n-1)!}{q!\,(n-q)!}\,\right)^{1/n}\\
&\le& \E^{1/\e}\ \frac{n}{aq^{q/n}\,(n-q)^{(n-q)/n}}.\label{above}
\eea
An explicit computation shows that
$$
\frac{\partial\phantom{n}}{\partial n}\ \,\frac{n}{q^{q/n}\,(n-q)^{(n-q)/n}}
\ =\ \frac{n}{q^{q/n}\,(n-q)^{(n-q)/n}}\
     \frac{q}{n^2}\ \left(\,\ln(q)-\ln(n-q)\,\right) .
$$
From this, we deduce that the right hand side of (\ref{above}) attains its
maximum at $n=2q$. Evaluating that maximum, we obtain
\be\label{half1}
\sup_{q<n\le [q(p+1)/2]}\ (G_{p}(n,\,q))^{1/n}\ \leq\ \frac{2}{a}\,\E^{1/\e}.
\ee

For $[q(p+1)/2]<n<pq$, the number $n'=q(p+1)-n$ satisifes
$q<n'\le [q(p+1)/2]$ and $n'/n\le 1$.
Thus, by (\ref{gsym}) and (\ref{half1}), we obtain
\bea\nonumber
(G_{p}(n,\,q))^{1/n}&=&(G_{p}(n',\,q))^{1/n}\\
\nonumber &\le&\left(\,\frac{2}{a}\,\E^{1/\e}\,\right)^{n'/n}\\
\nonumber &\le&\frac{2}{a}\,\E^{1/\e}.
\eea
Thus,
\be\label{half2}
\sup_{[q(p+1)/2]<n<pq}\ (G_{p}(n,q))^{1/n}\ \leq\ \frac{2}{a}\,\E^{1/\e}.
\ee

Inequalities (\ref{half1}) and (\ref{half2}) imply the
existence of $C_7$ for which (\ref{escom1}) holds.
This implies (\ref{escom}),and the lemma is proved. \ep

\vskip .25cm
We now generalize Lemma \ref{lem4} to the multi-dimensional case.

\vskip .25cm
\noindent
\begin{lem}\label{dlem4}
Let $F_p(n,\,q)$ be defined by (\ref{dcombin}).
For all $p\ge 1$, $q\ge 1$, and $n\ge 1$, we have
\be
F_p(n,q)\,\leq\,C_7^{(n+qd)},
\ee
where $C_7$ is the constant of Lemma \ref{lem4}.
\end{lem}

\vskip .25cm
\noindent
{\bf Proof:}\quad
With $m_j$ temporarily denoting numbers instead of multi-indices, we define
$$
\Gamma_p(n,\,q)=\sum_{\scriptsize\matrix{0\leq m_1,m_2,\cdots,m_q\leq p \cr
m_1+m_2+\cdots +m_q=n}}1,
$$
which is zero unless $0\leq n\leq pq$.

By defining $m'_j=m_j+1$, we see that
every decomposition of $n$ as $n=m_1+m_2+\cdots +m_q$, with $0\le m_j\le p$,
corresponds uniquely to a decomposition of $n+q$ as
$n+q=m'_1+m'_2+\cdots +m'_q$, with $1\le m'_j\le p+1$. Therefore,
we have the identity
\be\label{shift}
\Gamma_p(n,\,q)\,=\,G_{p+1}(n+q,\,q).
\ee

We now let the $m_j$ denote multi-indices and let $m_j(k)$ denote the
$k^{\mbox{\scriptsize th}}$ component of $m_j$.
Then, using (\ref{shift}) and Lemma \ref{lem4}, we easily obtain
$$
F_p(n,\,q)\ \leq \sum_{\scriptsize\matrix{0\leq m_j(k)\leq p \cr
\sum_{j=1}^q\sum_{k=1}^dm_j(k)=n}}1\ =\ \Gamma_p(n,\,qd)\ \leq\ C_7^{(qd+n)}.
$$
\phantom{}\ep

\vskip .25in
\noindent
{\bf Proof of Lemma \ref{truncest}:}\quad
Suppose $2\leq q \leq l+2$, and let $L_q(l)$ satisfy (\ref{Lest}), {\it i.e.},
$$
L_q(l)\,\le\,\frac{C_1^q}{\hbar^q\,q!}\,
\sum_{n=3l-2}^{q(l+1)}\,(C_2\,\hbar\,l)^{n/2}\,F_{l+1}(n,\,q).
$$
By Lemma \ref{dlem4} and (\ref{stirling}),
there exist $C_8$ and $C_9$, such that
\bea
L_q(l)&\leq&\frac{(C_1\,C_7^d)^q}{(a\,\hbar\,q)^q}\,
\sum_{n=3l-2}^{q(l+1)}\,\left(C_2\,\hbar\,l\right)^{n/2}\,C_7^{n}
\nonumber \\ \label{start}
&\leq&\frac{C_9^q}{(\hbar\,q)^q}\,
\sum_{n=3l-2}^{q(l+1)}\,\left(C_8\,\hbar^{1/2}\,l^{1/2}\right)^{n}.
\eea

Note that we can take
\bea\label{c8}
C_8&=&\max\,\{\,1,\,\sqrt{C_2}\,C_7\,\},\\ \label{c9}
C_9&=&\max\,\{\,1,\,C_1\,C_7^d\,\},
\eea
because we can assume $C_8\ge 1$ and $C_9\ge 1$ without loss of generality.

We arbitrarily choose two positive numbers $\alpha<1$ and $\beta<1$. We then
define
\be\label{gchoice}
g\,=\,\frac{1}{C_8^2}\,
\min\,\left\{\,\alpha^2,\,\frac{\beta^2}{C_9^2\,C_8^4\,\E^{2/\e}}\,\right\} .
\ee
We henceforth assume $l=l(\hbar)$ is chosen to satisfy
(\ref{lchoice}) with this value of $g$. With this choice, we have
\be\label{condal}
C_8\,\hbar^{1/2}\,l^{1/2}\,\leq\,\alpha .
\ee
By summing a geometric series in (\ref{start}), we see that
\bea
L_q(l)&\leq&\frac{C_9^q}{(1-\alpha)\,(\hbar\,q)^q}\
\left(\,C_8\,\hbar^{1/2}\,l^{1/2}\,\right)^{3l-2}\nonumber\\[5pt]
&=&\frac{C_9^q\,C_8^{2q}}{(1-\alpha)\,(\hbar\,q\,C_8^2)^q}\,
\left(\,C_8\,\hbar^{1/2}\,l^{1/2}\,\right)^{3l-2}.
\eea
As a function of $q$,
$$
\left(\frac{l}{q}\right)^q=\exp\left(\,q\,\ln(l/q)\,\right)
$$
is maximized at $q=l/\E$, where it has the value $\E^{l/\e}$.
Thus,
$$
\left(\frac{l}{q}\right)^q\,\le\,\E^{l/\e}.
$$
This, $q\le l+2$, and $\hbar\,l\,C_8^2\,\leq\,\alpha^2\,<\,1$
imply
$$
\frac{1}{(\hbar\,q\,C_8^2)^q}\ =\ \frac{1}{(\hbar\,l\,C_8^2)^q}\,
\left(\,\frac{l}{q}\,\right)^q
\ \leq\ \frac{\E^{l/\e}}{(\hbar\,l\,C_8^2)^{l+2}}.
$$
From this, $C_8\geq 1$, and $C_9\geq 1$, we conclude
\bea\label{etap}
L_q(l)&\leq&\frac{C_9^{l+2}\,C_8^{2(l+2)}\,\E^{l/\e}}{(1-\alpha)}\
\left(\,C_8\,\hbar^{1/2}\,l^{1/2}\,\right)^{l-6}\nonumber\\[5pt]
&=&\left(\,\E^{1/\e}\,C_9\,C_8^3\,\hbar^{1/2}\,l^{1/2}\,\right)^l\
\frac{C_9^2}{(1-\alpha)\,(\hbar\,l)^3\,C_8^2}.
\eea
By (\ref{gchoice}), we have
$$
\E^{1/e}\,C_9\,C_8^3\,\hbar^{1/2}\,l^{1/2}\ \leq\ \beta\ <\ 1.
$$
Since
$$
\frac g\hbar -1\ \le\ l\ \le\ \frac g\hbar ,
$$
we see from (\ref{etap}) that if $\hbar\leq g/2$, then we have
\bea
L_q(l)&\leq&\E^{-|\ln(\beta)|l}\
\frac{C_9^2}{(1-\alpha)\,(g-\hbar)^3\,C_8^2}\nonumber\\[5pt] \label{425}
&\leq&\frac{2^3\,\E\,C_9^2}{(1-\alpha)\,g^3\,C_8^2}\
\E^{-|\ln(\beta)\,|g/\hbar}.
\eea
Note that the factor $2^3$ can be replaced
by $(1+\epsilon)$, with $\epsilon$ arbitrarily small, by taking
$\hbar$ sufficiently small.\ep

\vskip .25cm \noindent
{\bf Remark}\quad
We can choose $\alpha$ and $\beta$ to satisfy
$\ds \frac{\beta}{\E^{1/\e}}\,\leq\,\alpha\,<\,\frac 1{\E^{1/\e}}$.
Then, since $\ds \frac{\beta^2}{C_9^2\,C_8^4\,e^{2/e}}\,\leq\,
\frac{\beta^2}{\E^{2/\e}}\,\leq\,\alpha^2$,
we obtain the conclusion of Lemma \ref{truncest} with
\bea\label{gchoice1}
g&=&\frac{\beta^2}{C_9^2\,C_8^6\,\E^{2/\e}},\\
\label{gamchoice}
\gamma_1&=&\frac{|\ln(\beta)|\,\beta^2}{C_9^2\,C_8^6\,e^{2/e}}\\
\label{Cchoice}
C_3&=&\frac{2^3\,\E^{1+6/\e}\,C_9^8\,C_8^{10}}{(1-\E^{-1/\e})\,\beta^6},
\eea
where $C_8$ and $C_9$ are related to $C_1$, $C_2$, and $C_7$
by (\ref{c8}) and (\ref{c9}).
Note that $C_7$ is purely combinatorial and has no time dependence.

\vskip .5cm
\section{Proof of Theorem 1.2} 
\setcounter{equation}{0}

To prove Theorem \ref{longtime}, we revisit the proof of Theorem
\ref{mainthm} and make the $T$ dependence of all constants explicit.
We then allow $T$ to grow with $\hbar$ with the restriction that our
approximation remain close to the actual solution as $\hbar\ra 0$.

Lemmas \ref{lem3}, \ref{lem4}, and \ref{dlem4} have no time
dependence, and the
time dependence of Lemmas \ref{fundcalc}, \ref{taylorprop},
\ref{dlem2},
\ref{taylor1}, \ref{dlem0}, and
\ref{dlem1} has been made explicit in their conclusions.

Since $V$ is bounded below and energy is conserved,
$|a(t)|$ grows at most linearly with
time. Thus, there exist $v_1 >0$ and $v_0>0$, such that
$v(t)$ defined by (\ref{mess88}) satisifies
\be\label{1}
v(t)\,\leq\,v_0\,\exp(v_1 t).
\ee
From assumption (\ref{expgro}), it follows that
\be\label{2}
w(T)\,=\,\sup_{0\leq |t|\leq T}\,\|A(t)\|\,\leq\,N\,\exp (\lambda T).
\ee

In the sequel, we denote all innessential constants that do not depend
on $T$ by the same symbol $c$.

Consider Lemma \ref{taylorprop}.
For $z\in C_{\delta}(\zeta(x,a(t)))$,
we can prove the existence of $\rho>0$ such that
\bea\nonumber
&&\left\|\,\frac{D^mV(\zeta(x,a(t)))}{m!}\ (x-a(t))^m\
\psi_{l}(x,\,t,\,\hbar)\,\right\|\\[5pt] \label{goal2}
&\le& c\,\exp(\rho |t|)\,
\left(c\,\sqrt{\hbar\,(l+2)}\,\exp (\lambda |t|)\right)^{l+2}.
\eea
Indeed, with our bound on $|V(z)|$ we have
$$\frac{D^mV(\zeta(x,a(t)))}{m!}\,\leq\,c\,\exp(\rho |t|)\,\exp (c|x-a(t)|)
$$ instead of (\ref{mess8}).
With this and the estimate $\exp(cy)\leq\exp(cy^2)+\exp (c)$ for all
$c>0$ and $y>0$, we see that
the estimates in the proof of Lemma \ref{taylorprop} are still valid.
This yields (\ref{goal2}).
Consequently, the
constants $C_4$ and $C_5$ appearing in (\ref{taylordone}) satisfy
\bea\nonumber
C_4(T)&=&c\,\exp(\rho T)\\ \nonumber
C_5(T)&=&c\,\exp(\lambda T),
\eea
where $\rho>0$.
By using estimates (\ref{1}) and (\ref{2}) in (\ref{cqest1})
and (\ref{rest1}), we see that when we apply Lemma \ref{truncest},
the constants $C_1$ and $C_2$ satisfy
\bea\nonumber
C_1(T)&=&c\,\exp(\rho T)\\ \nonumber
C_2(T)&=&c\,\exp(\lambda T).
\eea
We still must determine the $T$ dependence of $C$ and $\gamma$ of
Theorem \ref{mainthm}. We do this
by determining the $T$ dependence of $g, \gamma_1$ and
$C_3$ in Lemma \ref{truncest}. Equations
(\ref{c8}) and (\ref{c9}) yield $C_8$ and $C_9$, in terms of which the
above constants are determined.
We find that
\bea\nonumber
C_8(T)&=&c\,\exp(\lambda T/2)\\ \nonumber
C_9(T)&=&c\,\exp(\rho T).
\eea
Consequently, from (\ref{gchoice1}), (\ref{gamchoice}) ,(\ref{Cchoice}),
we have
\bea\nonumber
g(T)&=&c\,\exp(-\nu T),\qquad\qquad\nu=2\rho+3\lambda\\ \nonumber
\gamma_1(T)&=&c\,\exp(-\nu T),\\ \nonumber
C_3(T)&=&c\,\exp(\mu T),\qquad\qquad\ \,\mu=8\rho+5\lambda .
\eea

To arrive at these conclusions, we imposed various conditions.
Those conditions were\\[3pt]
\phantom{}\qquad\qquad\qquad
(\ref{35}), that now requires $\hbar\leq c \exp(-2\lambda T)$,\\[3pt]
\phantom{}\qquad\qquad\qquad
(\ref{george}), than now requires $c\exp(2\lambda T)\exp(-\nu T) <1$,
and\\[3pt]
\phantom{}\qquad\qquad\qquad
$ \hbar\leq g(T)/2$ used in (\ref{425}),
that now requires $ \hbar\leq c\exp(-\nu T)$.

\vskip .15cm
These are all satisfied, provided we take
$$
  T(\hbar)\,\leq\,T'\,|\ln(\hbar)|,
$$
with $T'>0$ sufficiently small. Using this in our estimate for the
error term $\xi(t,\hbar)$, we see that for sufficiently small $T'$,
$|t|\,\leq\,T'\,|\ln(\hbar)|$ implies
$$
\| \psi(x,t,\hbar)-\Psi(x,t,\hbar)\|_{L^2(\r^d)}\,\leq\,
C'\,\exp(-\gamma'/\hbar^{\sigma}),
$$
for some $\gamma'>0$, $\sigma >0$, and $C'>0$.\ep

\end{document}